\documentclass[a4paper,10pt]{article}
\usepackage[utf8]{inputenc}
\usepackage{amsthm}
\usepackage{amsfonts}
\usepackage{amssymb}
\usepackage{amsmath}
\usepackage{bbold}
\allowdisplaybreaks
\usepackage{mathtools}
\usepackage[english]{babel}
\usepackage{color}
\usepackage{slashed}
\usepackage{enumerate}
\usepackage{graphicx}
\usepackage{systeme}
\usepackage{dsfont}
\usepackage{relsize}
\usepackage[margin=0.90in]{geometry}
\usepackage{float}
\usepackage{siunitx}
\usepackage[labelformat=simple]{subcaption}

\usepackage{graphicx}
\usepackage{bmpsize}
\usepackage{epstopdf}
\usepackage{bm}
\usepackage{bbm}
\usepackage[dvipsnames]{xcolor}
\usepackage{etoolbox}
\usepackage{titling}
\thanksmarkseries{arabic} 
\usepackage{authblk}

\usepackage{blindtext,graphicx}
\usepackage[absolute]{textpos}
\usepackage{physics}
\AtBeginDocument{\RenewCommandCopy\qty\SI}
\ExplSyntaxOn
\msg_redirect_name:nnn { siunitx } { physics-pkg } { none }
\ExplSyntaxOff
\usepackage[hang,flushmargin]{footmisc}

\usepackage{xfrac}
\usepackage{nicefrac}
\usepackage{esvect}
\usepackage{cases}
\usepackage{empheq}
\usepackage{stmaryrd}
\usepackage{cancel}
\usepackage[linguistics]{forest}
\usepackage{url}
\usepackage[font=small]{caption}
\usepackage{multirow}
\newcommand{\comsol}{\textit{Comsol Multiphysics}\textsuperscript{\textregistered}}

\usepackage[section]{placeins}
\usepackage[percent]{overpic}
\usepackage{tikz}

\usepackage{setspace}
\usepackage{forest}

\usepackage[giveninits=true,sortcites=true,date=year,maxbibnames=99,doi=false,isbn=false,url=false,eprint=false]{biblatex}
\renewbibmacro{in:}{}
\DeclareFieldFormat{pages}{#1}
\AtEveryBibitem{%
  \clearlist{language}%
}
\usepackage{csquotes}

\usepackage{setspace}

\usepackage{tikz}
\usetikzlibrary{decorations.pathmorphing, patterns}
\usetikzlibrary{shapes.geometric, arrows.meta, positioning}

\tikzstyle{decision} = [
    diamond,
    draw,
    text width=4cm,
    text badly centered,
    inner sep=1pt
]

\tikzstyle{arrow} = [thick,->,>=Stealth]
\title{Multiple Band-Gaps through the Coupling of Unit Cells from the Same Metamaterial: the Dual Cell method}
\author{
  Plastiras Demetriou\thanks{Faculty of Architecture and Civil Engineering, TU Dortmund, August-Schmidt-Str. 8, 44227 Dortmund, Germany}
  \quad
}

\thanksmarkseries{arabic}
\date{\today}
\begin{document}
\maketitle

%
%
%
%
\begin{abstract}
This study investigates how the coupling of two unit cells belonging to the same mechanical metamaterial into a dual unit cell configuration, can produce a new metamaterial with enhanced wave attenuation capabilities.
For two metamaterials, two different unit cell coupling configurations are examined in 2D (side by side and chessboard), with particular emphasis on maintaining a plane crystallographic group of high symmetry, in order to simplify band structure calculations given the complexity of the geometry.
It is shown that for specific configurations and choices of unit cell, multiple directional and/or omnidirectional band-gaps can appear, some of which can exhibit enhanced attenuation.
The way in which these band-gaps emerge is described through applying the same procedure on 1D spring mass chains.
Results support the idea that any band-gap metamaterial could have a much more efficient version which can be constructed purely from its own unit cells. 
 
\end{abstract}

%
%
%
%
\section{Introduction}
\label{sec:intro}
In recent years, mechanical metamaterials have gained increasing interest due to their capability of controlling the propagation of mechanical waves in exotic ways. 
Their unusual dynamic behavior arises both from the properties of their base materials and from the design of their microstructure.
If this design is done appropriately, the metamaterial can exhibit band-gaps, which are frequency regions where propagating waves are effectively attenuated.
Many applications are therefore possible, including vibration isolation and wave shielding \cite{al2022advances, miniaci2016large, otlu2023three} and energy harvesting \cite{de2021graded, li2017design, de2020experimental}.
Mechanical metamaterials can also offer other advanced functionalities such as blast and impact protection \cite{vo2022blast, haid2023mechanical, fisher2025ballistic}.

Metamaterials with band-gaps are usually periodic.
As a result of this periodicity, the design of the whole structure collapes to the design of its structural unit, the unit cell.
Depending on the design of the unit cell, band-gaps may arise from different physical mechanisms, of which the three most known are Bragg Scattering \cite{brillouin1953wave}, Local Resonance \cite{liu2000locally, al2017formation} and Inertial Amplification \cite{yilmaz2007phononic}. 
Bragg Scattering can produce band-gaps due to destructive interference of the waves, with band-gaps appearing at frequencies related to the lattice periodicity.
The lowest Bragg gap can appear at frequencies of the order of the wave speed of the base material divided by the lattice constant \cite{goffaux2003two, yilmaz2007phononic}.
For Local Resonance, it is possible that a band-gap opens at lower fequencies than Bragg Scattering since the opening is associated to the eigenfrequency of the resonators, but such band-gaps are usually narrow \cite{yilmaz2007phononic}.
To overcome this limitation, design methodologies with multiple resonators have been proposed \cite{tian2019elastic, zhao2021multi, roca2021multiresonant}, some of which employ the so-called dual unit cell which has been introduced in \cite{gao2020ultrawide} and studied in great detail in \cite{stein2022widening} \footnote{although the authors of \cite{gao2020ultrawide} used the term ``two simple cells (TCS) hybrid system", the simpler term ``dual" was used later in \cite{stein2022widening} to refer to that specific dual-periodic system.}.
Additionally, Inertial Amplification has been proposed specifically for creating wide low-frequency band-gaps by essentially using inerters \cite{smith2002synthesis} in a periodic arrangement \cite{yilmaz2017dynamics, yilmaz2007phononic, otlu2023three}.
Finally, designs have been proposed that take advantage of two or even all three discussed band-gap mechanisms \cite{yuan2013coupling, krushynska2017coupling, mazzotti2023bio, li2023seismic}.

In order to identify band-gaps, the dispersion behavior is commonly evaluated through a Bloch-Floquet analysis \cite{cool2024guide} on a unit cell of minimum volume (i.e. a primitive unit cell), which allows to produce the band structure (dispersion diagrams/ dispersion plot) of the metamaterial.
Such analysis can be performed on any primitive unit cell of the metamaterial since the Bloch-Floquet theorem concerns infinite periodic systems, implying that the choice of unit cell is not unique \cite{demetriou2024reduced, cool2022impact, cool2024guide}. 

The main goal of this paper is to investigate the properties of metamaterials with unit cells made out of the coupling of two different unit cells from an already existing metamaterial design.  
This concept of combining unit cells from the same metamaterial, we call the dual cell method.
The term dual cell was used in \cite{stein2022widening} to refer in particular to resonators assembled in a super cell dual periodic configuration (i.e. two (generally) different locally resonant cells connected in series). 
In this work, the term dual cell represents an assembly of any two unit cells as long as they belong to the same initial metamaterial, and not necessarily belonging to a locally-resonant metamaterial, but rather, to any band-gap metamaterial. 
The dual cell method is demonstrated first on 1D infinite spring-mass chains where it is shown that it produces chains with additional band-gaps, some of which also exhibit enhanced attenuation. 
The method is then applied to 2D metamaterials, where two different ways of coupling unit cells are shown in which multiple band-gaps emerge.

\textbf{The article is structured as follows:} section~\ref{sec:intro} introduces the topic and the motivation; section~\ref{sec:cutNshape} briefly describes the choice of unit cell in periodic metamaterials; section~\ref{sec:unit_cells} presents the parent metamaterials of this study and associates an infinite chain to each one given the band-gap mechanism responsible for the band-gap that has been engineered in each; section ~\ref{sec:dual_cell_method} defines the dual cell method and applies it to the 1D infinite chains of section~\ref{sec:unit_cells}; section ~\ref{sec:dual_2D} presents the dual cell method in 2D, by applying it to the parent metamaterials of section~\ref{sec:unit_cells} using two different unit cell coupling configurations and also presents a finite-sized test for the best performing metamaterial for Transmissibility evaluation; section~\ref{sec:concl} presents the conclusions of this study together with future perspective.
%
%
%
%
\section{The choice of Unit Cell}
\label{sec:cutNshape}
An infinitely extended periodic metamaterial can be constructed by the periodic repetition of a unit cell in space.
More strictly, in two (three)  dimensions, this means that the translation of a unit cell along integer multiples of two (three) linearly independent vectors covers the whole 2D plane (3D space) without gaps or overlaps.
Such unit cell is not unique in shape, position on the lattice, or even volume; as both primitive (of the smallest volume) and conventional unit cells (of bigger volume) can exist \cite{Ashcroft1976}.

Limiting ourselves to primitive unit cells, which are necessary for a correct Bloch-Floquet analysis in periodic metamaterials \cite{mukherjee2015phononic}, the process of choosing a primitive unit cell in 2D can be broken down in two steps \cite{demetriou2026transmissibility}: (\textit{i}) choosing a valid unit cell shape of minimum volume that can tile the plane and (\textit{ii}) choosing a ``cut" of that shape (i.e. choose the position from which the unit cell will be cut out from the lattice).

In this work, the shape of the initial unit cells before the coupling will always be square and only the cut will vary, producing different choices of a square-shaped unit cell.
After the coupling of two different unit cells, dual unit cells are formed which can be of rectangular or square shape.
In regards to their symmetry, we characterise it using the full crystallographic notation, following \cite{maurin2018probability}.
%
%
%
%
\section{Metamaterials, band structures and band-gap mechanisms}\label{sec:unit_cells}
In this section, the two ``parent" metamaterials are introduced.
The term parent is used to signal the metamaterial from which two different unit cells will be used as building blocks for creating new metamaterials.
The band-gap mechanisms responsible for the opening of band-gaps in each of the two parent metamaterials are explained first using 1-D spring mass chains and then identified through the imaginary part of the dispersion plot of the metamaterials. 
Associating spring-mass chains to each metamaterial facilitates the description of the dual cell method in the next section.
Dimensions and Bloch- Floquet band structures (dispersion curves) are given for each metamaterial, while the base material is the same for both: structural steel.
The material properties of structural steel can be seen in Table~\ref{tab:steel_properties}.
For all metamaterials and infinite spring-mass chains in this study, wave attenuation inside a band-gap at a specific frequency
is evaluated by the smallest value of the imaginary part of the wavevector at that frequency, similarly to \cite{nadejde2025mechanisms}, which indicates the exponential decay of the wave.

\begin{table}[h]
	\centering
	\begin{tabular}{|l|c|c|}
		\hline
		\textbf{Property} & \textbf{Symbol} & \textbf{Value} \\
		\hline
		Elastic Modulus & $E$ & \SI{210}{\giga\pascal} \\
		\hline
		Poisson's Ratio       & $\nu$ & 0.30 \\
		\hline
		Density               & $\rho$ & \SI{7850}{\kilogram\per\cubic\meter} \\
		\hline
	\end{tabular}
	\caption{Properties of Structural Steel}
	\label{tab:steel_properties}
\end{table}

\subsection{Bragg Scattering: Square-Circular-Hole (SCH) metamaterial}
\label{sec:SCH}
One simple way that Bragg Scattering can be realised in 1-D is by using the alternating spring-mass chain with two different masses (i.e. the mechanical diatomic chain) \cite{brillouin1953wave,yilmaz2017dynamics}, thus representing the periodic variation of mass.
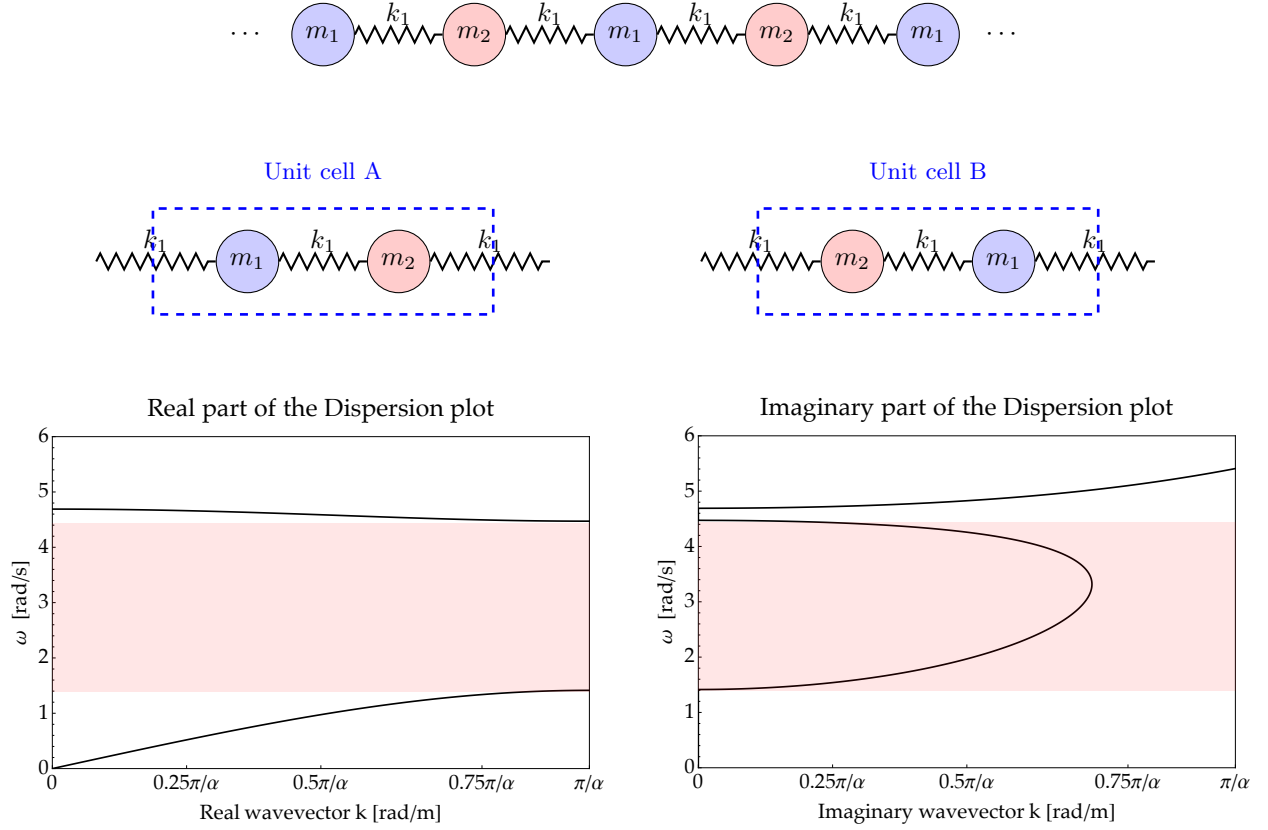
\begin{figure}[h]
	\centering 
	\begin{tikzpicture}[
		spring/.style={thick,decorate,
			decoration={zigzag,segment length=6,amplitude=3}},
		mass1/.style={circle,draw,fill=blue!20,minimum size=8mm},
		mass2/.style={circle,draw,fill=red!20,minimum size=8mm}
		]
		
		\node at (2,0) {$\cdots$};
		
		\node[mass1] (a1) at (3,0) {$m_1$};
		\node[mass2] (a2) at (5,0) {$m_2$};
		\node[mass1] (a3) at (7,0) {$m_1$};
		\node[mass2] (a4) at (9,0) {$m_2$};
		\node[mass1] (a5) at (11,0) {$m_1$};
		
		\draw[spring] (a1)-- node[above]{$k_{1}$} (a2);
		\draw[spring] (a2)-- node[above]{$k_{1}$} (a3);
		\draw[spring] (a3)-- node[above]{$k_{1}$} (a4);
		\draw[spring] (a4)-- node[above]{$k_{1}$} (a5);
		
		\node at (12,0) {$\cdots$};
		
		\node[blue] at (3,-1.8) {\small Unit cell A};
		
		\coordinate (s1) at (0,-3);
		\node[mass1] (u1) at (2,-3) {$m_1$};
		\node[mass2] (u2) at (4,-3) {$m_2$};
		\coordinate (s2) at (6,-3);
		
		\draw[spring] (s1)--node[above]{$k_{1}$}(u1);
		\draw[spring] (u1)--node[above]{$k_{1}$}(u2);
		\draw[spring] (u2)--node[above]{$k_{1}$}(s2);
		
		\draw[blue, line width=1pt, dashed] (0.75,-3.7) rectangle (5.25,-2.3);
		
		\node[blue] at (11,-1.8) {\small Unit cell B};
		
		\coordinate (t1) at (8,-3);
		\node[mass2] (v1) at (10,-3) {$m_2$};
		\node[mass1] (v2) at (12,-3) {$m_1$};
		\coordinate (t2) at (14,-3);
		
		\draw[spring] (t1)--node[above]{$k_{1}$}(v1);
		\draw[spring] (v1)--node[above]{$k_{1}$}(v2);
		\draw[spring] (v2)--node[above]{$k_{1}$}(t2);
		
		\draw[blue, line width=1pt, dashed] (8.75,-3.7) rectangle (13.25,-2.3);
		
	\end{tikzpicture}
	\vspace{1cm} 
	
	
	\begin{subfigure}[b]{0.48\textwidth}
		\centering
		\includegraphics[width=\textwidth]{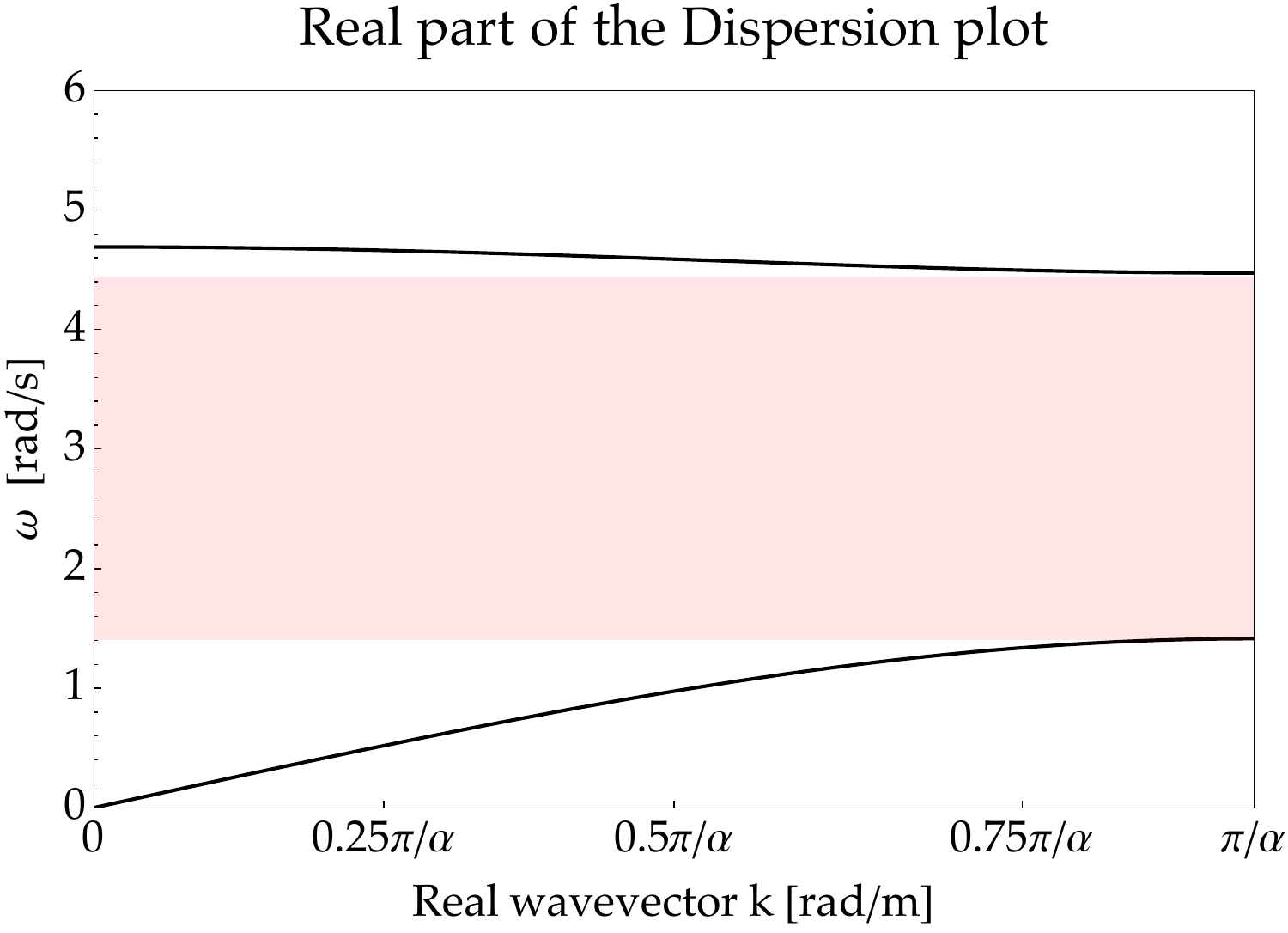}
	\end{subfigure}
	\hfill 
	\begin{subfigure}[b]{0.48\textwidth}
		\centering
		\includegraphics[width=\textwidth]{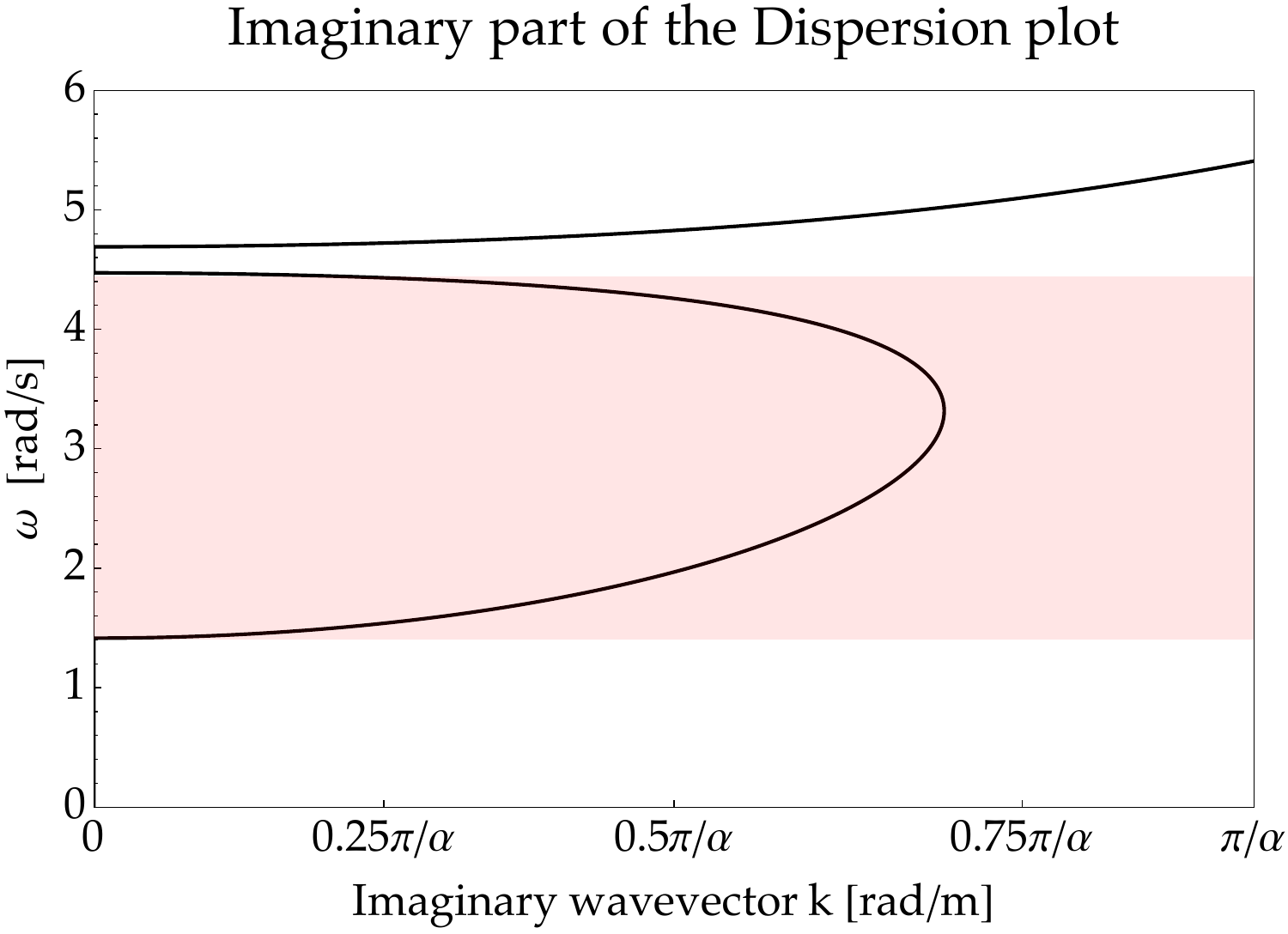}
	\end{subfigure}
	\caption{(Top) infinite diatomic chain and  two different choices of unit cell (bottom) Dispersion plot for the diatomic chain with (left) real part and (right) imaginary part. The Bragg gap is highlighted in red.}
	\label{fig:diatomic}
\end{figure}
This chain, together with two possible unit cells related by a half-period shift can be seen in Fig.~\ref{fig:diatomic}(top) with corresponding dispersion curves of the chain on the bottom.
For the parameter values and equilibrium equations in matrix form see Appendix \ref{sec:eq_eq}.
Since this is a two degree of freedom system, it produces two dispersion curves and there is a possibility of a band-gap when $m1 \neq m2$.
In the imaginary part of the plot we observe the characteristic Bragg Scattering attenuation profile, where attenuation varies smoothly through the gap with the maximum attenuation around the mid-gap frequency \cite{krushynska2017coupling, yilmaz2017dynamics, nadejde2025mechanisms}.

\begin{figure}[H]
	\centering
	\includegraphics[width=0.4\textwidth]{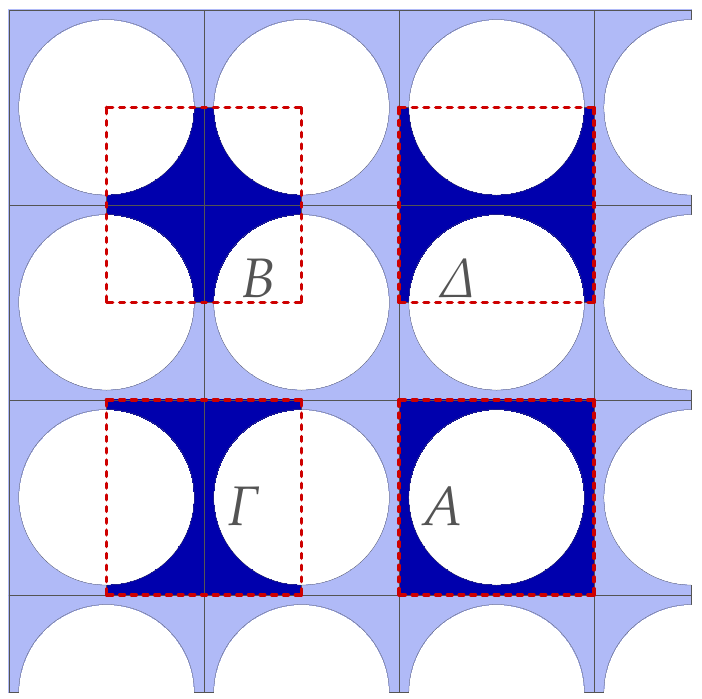}
	\includegraphics[width=0.4\textwidth]{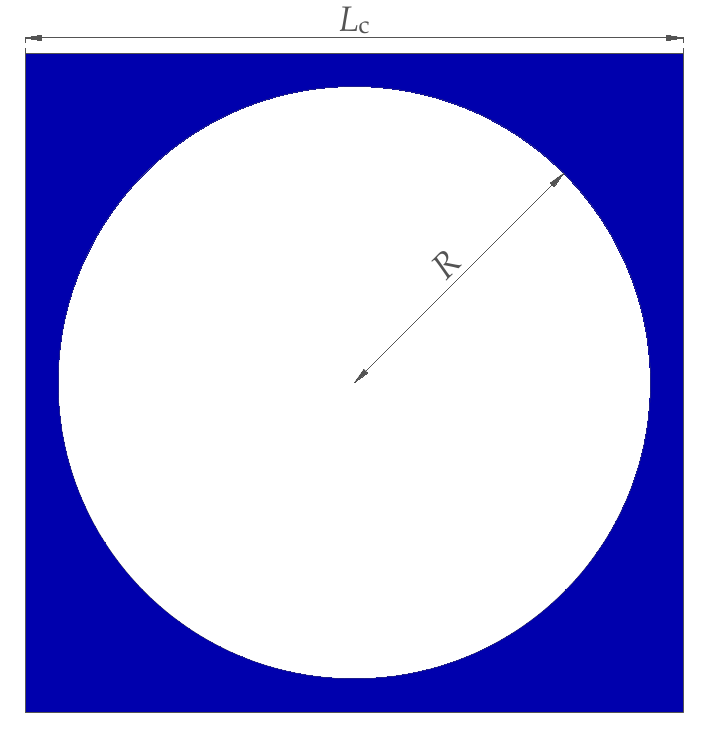}
	\vspace{0.3cm} 
	
	\begin{tabular}{|c|c|}
		\hline
		\textbf{Parameter} & \textbf{Value} \\
		\hline\
		$L_{c}$ & \SI{50}{\milli\meter}\\
		\hline
		$R$ & $\SI{22.5}{\milli\meter}$ \\
		\hline
	\end{tabular}
	\caption{(Left) the SCH metamaterial with four choices of primitive unit cell ($A, B, \Gamma$ and $\Delta$) and (right and bottom) geometrical dimensions shown on unit cell $A$.}
	\label{fig:SCH_unit_cell}
\end{figure}

\begin{figure}[!h]
	\centering
	\includegraphics[width=0.48\textwidth]{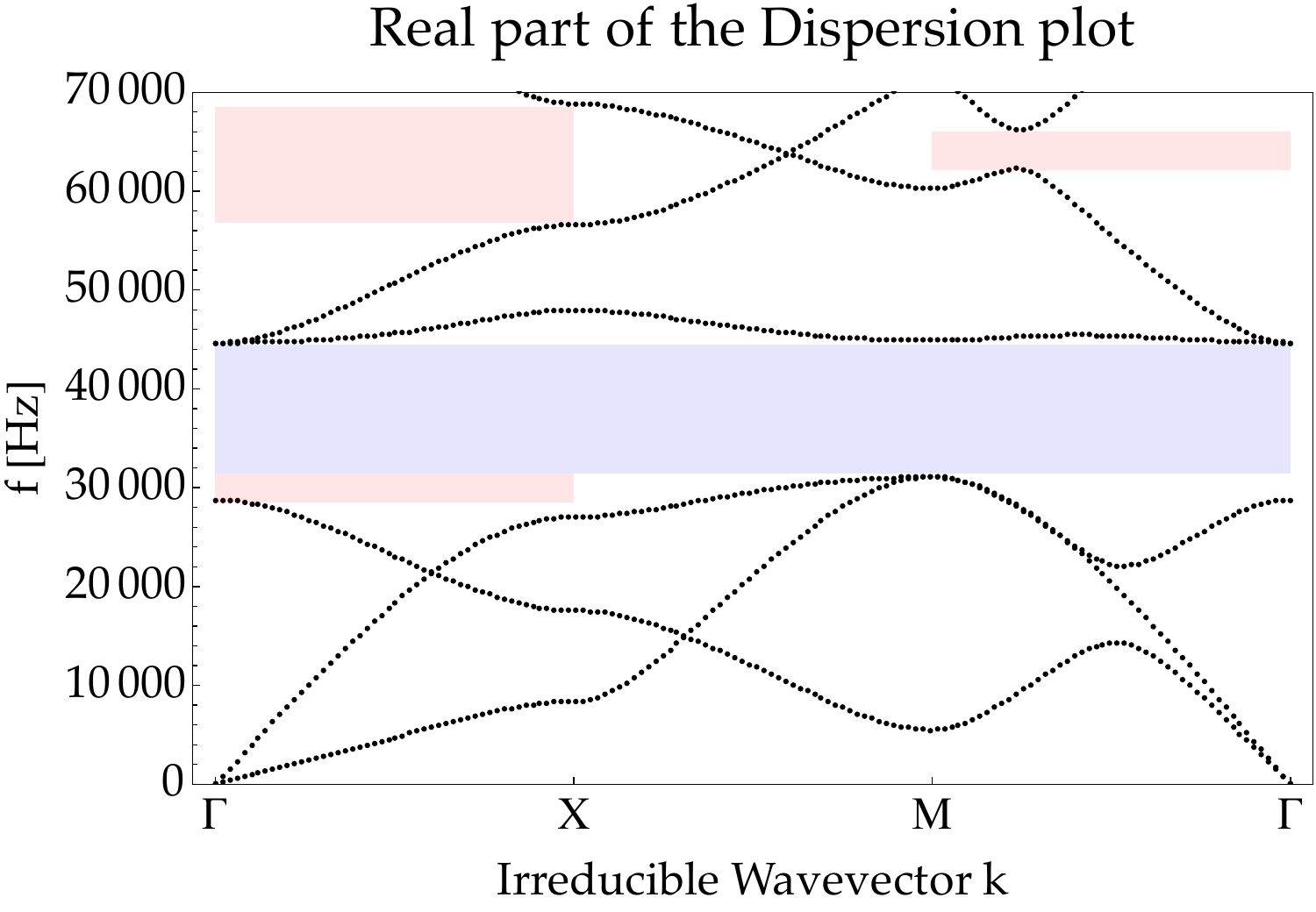}
	\includegraphics[width=0.48\textwidth]{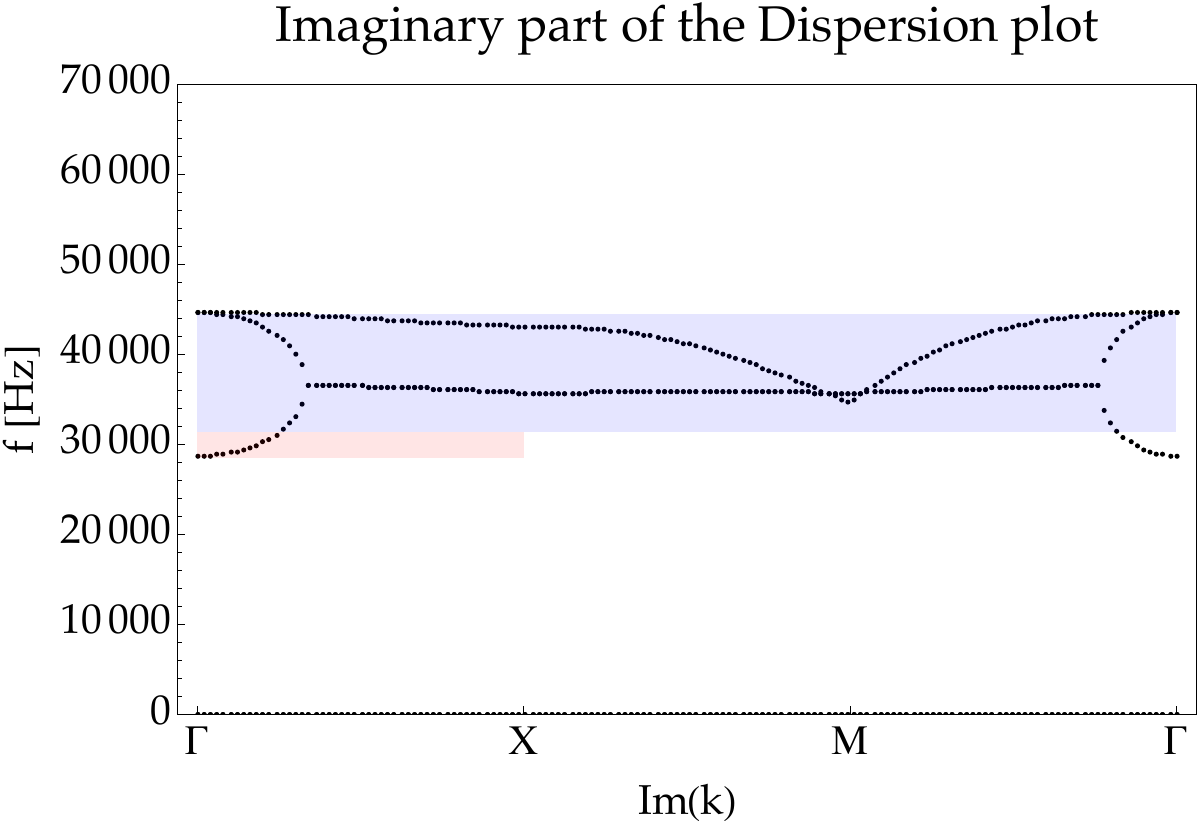}
	\caption{ Band Structure for the SCH metamaterial with (left) real part and (right) imaginary part. Omnidirectional band-gaps are highlighted in blue while directional band-gaps in red.}
	\label{fig:SCH_unit_cell_DC}
\end{figure}

The Square-Circular-Hole (SCH) metamaterial (see Fig.~\ref{fig:SCH_unit_cell}) is a very simple design which shows a band-gap through Bragg Scattering and as the name states, can be constructed by drilling a circular hole in a square. 
This metamaterial design belongs to the plane crystallographic group $p4mm$.
Four choices of unit cell colored with dark blue and inside red dashed-line squares, together with dimensions, can be seen in Fig.~\ref{fig:SCH_unit_cell}. 
The band structure can be seen in Fig.~\ref{fig:SCH_unit_cell_DC}, where a band-gap can be identified in the real part of the plot.
It is evident from the imaginary part of the plot that the band-gap mechanism responsible for the opening of the gap is Bragg Scattering, due to the attenuation profile. 

\subsection{Local Resonance: Four-Resonator metamaterial}
\label{sec:FR}
\begin{figure}[htbp]
	\centering
	\resizebox{0.9\textwidth}{!}{ 
		\begin{tikzpicture}[
			spring/.style={
				thick,
				decorate,
				decoration={zigzag, segment length=9, amplitude=4.5}
			},
			outerMass/.style={draw=black, line width=2.5pt},
			innerMass/.style={fill=red!20, draw=black, line width=1pt},
			cellBox/.style={draw=blue!70, dashed, line width=1pt, rounded corners=2pt}
			]
			
			\def\radiusOuterA{1.05}
			\def\radiusInnerA{0.33}
			
			\def\xA{0}
			\def\xB{3.9}
			\def\xC{7.8}
			\def\xD{11.7}
			
			\node[scale=1.5] at (-2.2, 0) {\Large $\dots$};
			\draw[spring] (-1.8, 0) -- (-\radiusOuterA, 0);
			\node[above] at ({(-1.8 + (-\radiusOuterA))/2}, 0.2) {$K_{1}$};
			
			\draw[outerMass] (\xA,0) circle (\radiusOuterA);
			\node[above] at (\xA, \radiusOuterA+0.2) {\large $M$};
			\draw[spring] (\xA-\radiusOuterA+0.01,0) -- (\xA-\radiusInnerA,0);
			\draw[innerMass] (\xA,0) circle (\radiusInnerA);
			\node[below] at (\xA, -\radiusInnerA-0.1) {$m$};
			\node[below] at (\xA-\radiusOuterA*0.6, -0.2) {$k_{1}$};
			
			\draw[spring] (\xA+\radiusOuterA,0) -- (\xB-\radiusOuterA,0);
			\node[above] at ({(\xA+\radiusOuterA+\xB-\radiusOuterA)/2}, 0.2) {$K_{1}$};
			
			\draw[outerMass] (\xB,0) circle (\radiusOuterA);
			\node[above] at (\xB, \radiusOuterA+0.2) {\large $M$};
			\draw[spring] (\xB-\radiusOuterA+0.01,0) -- (\xB-\radiusInnerA,0);
			\draw[innerMass] (\xB,0) circle (\radiusInnerA);
			\node[below] at (\xB, -\radiusInnerA-0.1) {$m$};
			\node[below] at (\xB-\radiusOuterA*0.6, -0.2) {$k_{1}$};
			
			\draw[spring] (\xB+\radiusOuterA,0) -- (\xC-\radiusOuterA,0);
			\node[above] at ({(\xB+\radiusOuterA+\xC-\radiusOuterA)/2}, 0.2) {$K_{1}$};
			
			\draw[outerMass] (\xC,0) circle (\radiusOuterA);
			\node[above] at (\xC, \radiusOuterA+0.2) {\large $M$};
			\draw[spring] (\xC-\radiusOuterA+0.01,0) -- (\xC-\radiusInnerA,0);
			\draw[innerMass] (\xC,0) circle (\radiusInnerA);
			\node[below] at (\xC, -\radiusInnerA-0.1) {$m$};
			\node[below] at (\xC-\radiusOuterA*0.6, -0.2) {$k_{1}$};
			
			\draw[spring] (\xC+\radiusOuterA,0) -- (\xD-\radiusOuterA,0);
			\node[above] at ({(\xC+\radiusOuterA+\xD-\radiusOuterA)/2}, 0.2) {$K_{1}$};
			
			\draw[outerMass] (\xD,0) circle (\radiusOuterA);
			\node[above] at (\xD, \radiusOuterA+0.2) {\large $M$};
			\draw[spring] (\xD-\radiusOuterA+0.01,0) -- (\xD-\radiusInnerA,0);
			\draw[innerMass] (\xD,0) circle (\radiusInnerA);
			\node[below] at (\xD, -\radiusInnerA-0.1) {$m$};
			\node[below] at (\xD-\radiusOuterA*0.6, -0.2) {$k_{1}$};
			
			\draw[spring] (\xD+\radiusOuterA,0) -- (\xD+\radiusOuterA+1.2,0);
			\node[above] at (\xD+\radiusOuterA+0.6, 0.2) {$K_{1}$};
			\node[scale=1.5] at (\xD+\radiusOuterA+1.7, 0) {\Large $\dots$};
			
			\draw[cellBox]
			(\xA+\radiusOuterA-1.05, -1.3)
			rectangle
			(\xB+\radiusOuterA-1.05,  1.6);
			\node[blue!70, below] at ({(\xA+\xB)/2+\radiusOuterA-1.05}, -1.3)
			{Unit Cell A};
			
			\draw[cellBox]
			(\xA+\radiusOuterA+4.8, -1.3)
			rectangle
			(\xB+\radiusOuterA+4.8,  1.6);
			\node[blue!70, below] at ({(\xA+\xB)/2+\radiusOuterA+4.8}, -1.3)
			{Unit Cell B};
		\end{tikzpicture}
	}
	\vspace{1cm} 
	
	
	\begin{subfigure}[b]{0.48\textwidth}
		\centering
		\includegraphics[width=\textwidth]{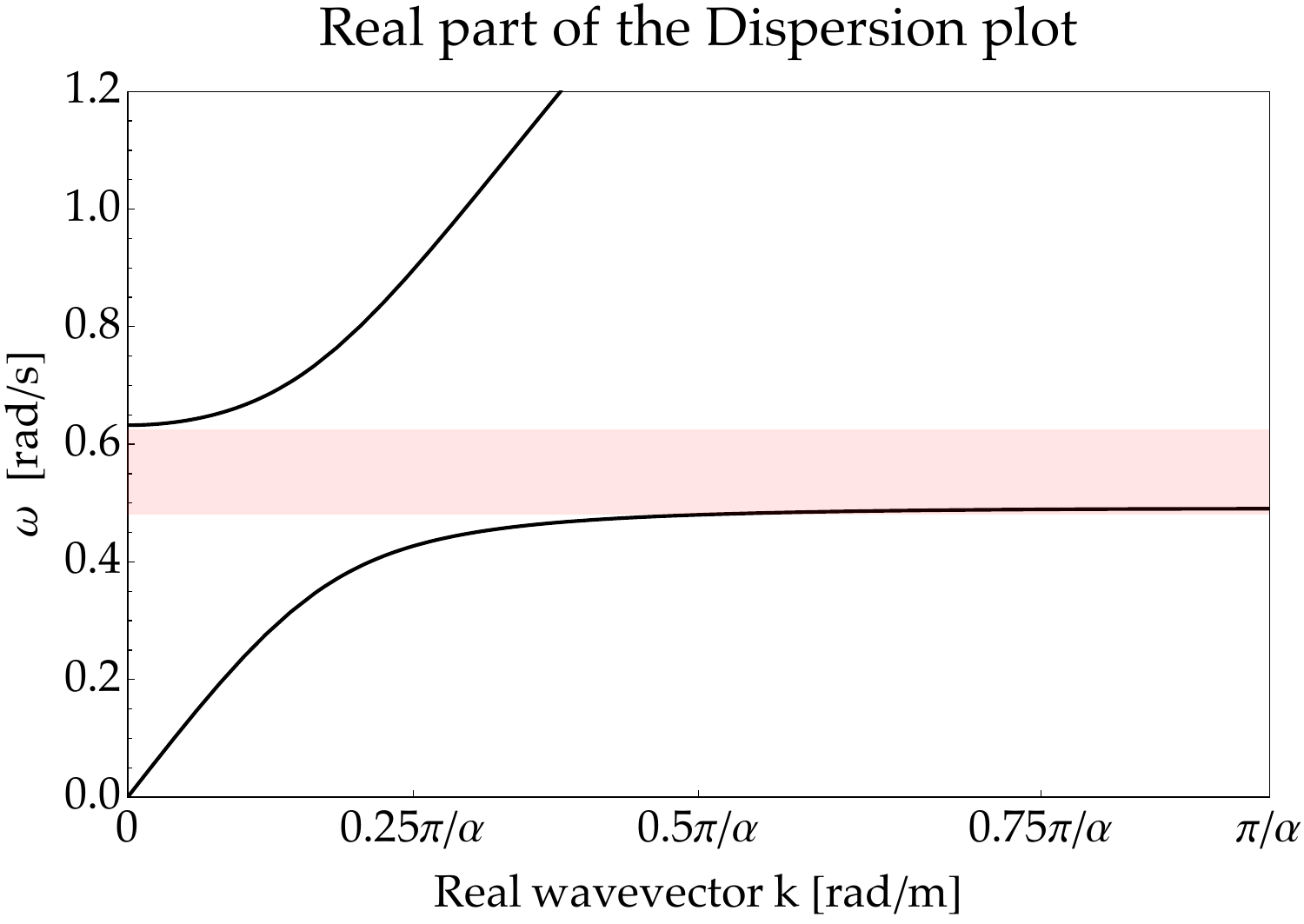}
	\end{subfigure}
	\hfill 
	\begin{subfigure}[b]{0.48\textwidth}
		\centering
		\includegraphics[width=\textwidth]{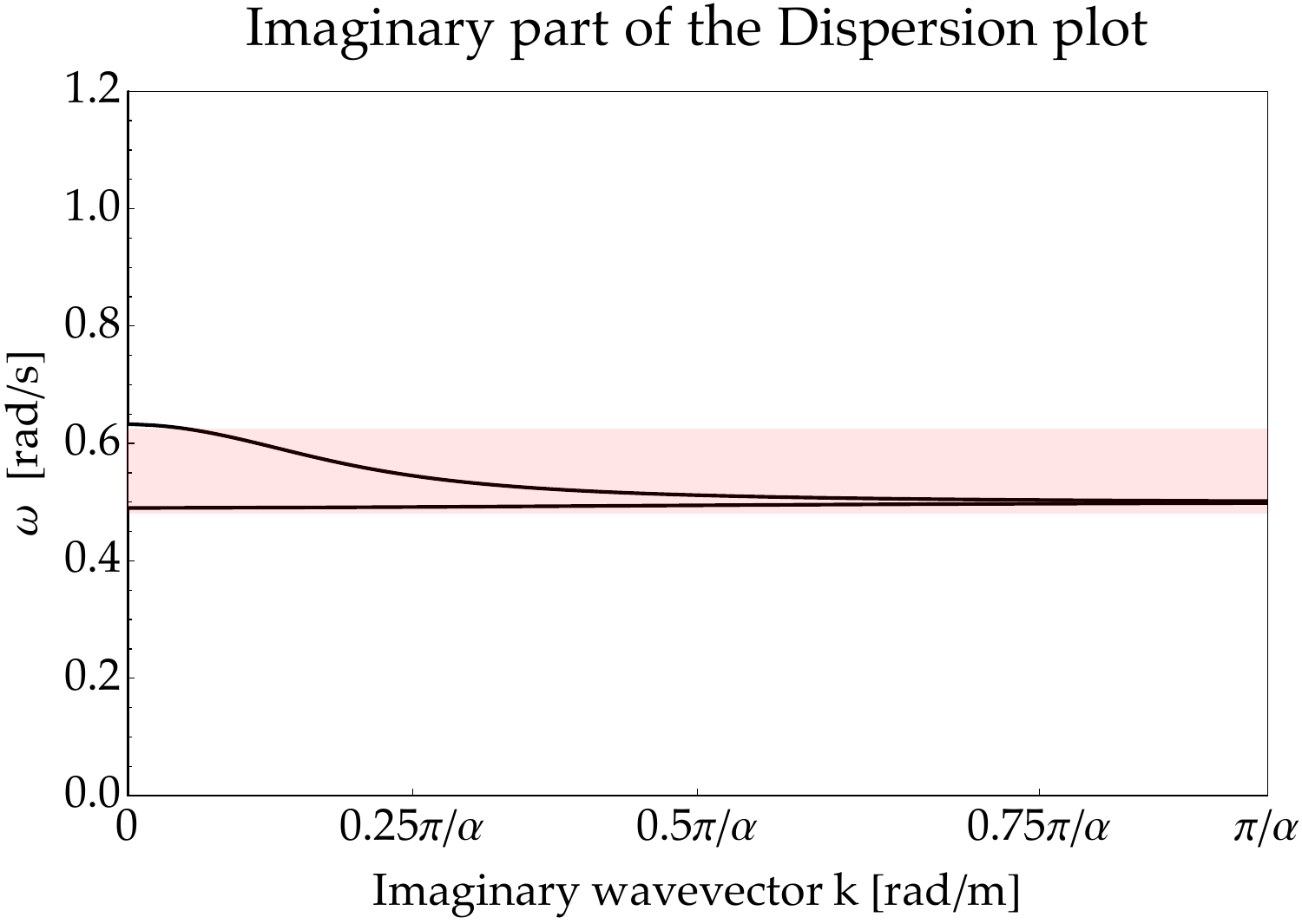}
	\end{subfigure}
	\caption{(Top) infinite mass-in-mass (Locally Resonant) chain and  two different choices of unit cell. (Bottom) dispersion plot for the mass-in-mass chain with (left) real part and (right) imaginary part. The Local Resonance band-gap is highlighted in red.}
	\label{fig:mass_in_mass}
\end{figure}
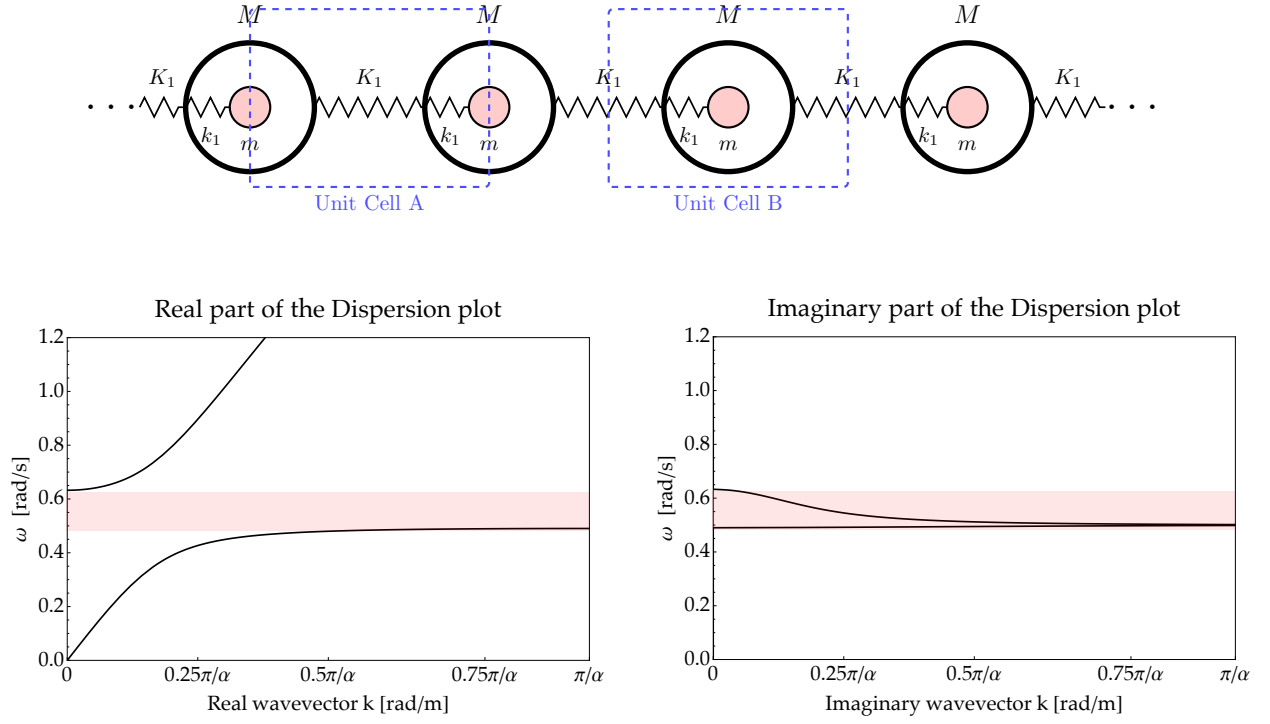

Local resonance can be realised in 1-D using the mass-in-mass infinite chain \cite{al2017formation,yilmaz2017dynamics} which can be seen in Fig.~\ref{fig:mass_in_mass}. 
Two different choices of unit cell are shown, which are related by a half-period shift.
The corresponding dispersion plot can be seen in Fig. \ref{fig:mass_in_mass} where we observe a narrow band-gap with the characteristic attenuation profile of Local Resonance: sharp attenuation around the eigenfrequency of the resonator \cite{krushynska2017coupling, yilmaz2017dynamics, nadejde2025mechanisms}.
For the parameter values and equilibrium equations in matrix form see Appendix \ref{sec:eq_eq}.

\begin{figure}[!htbp]
	\centering
	\includegraphics[width=0.4\textwidth]{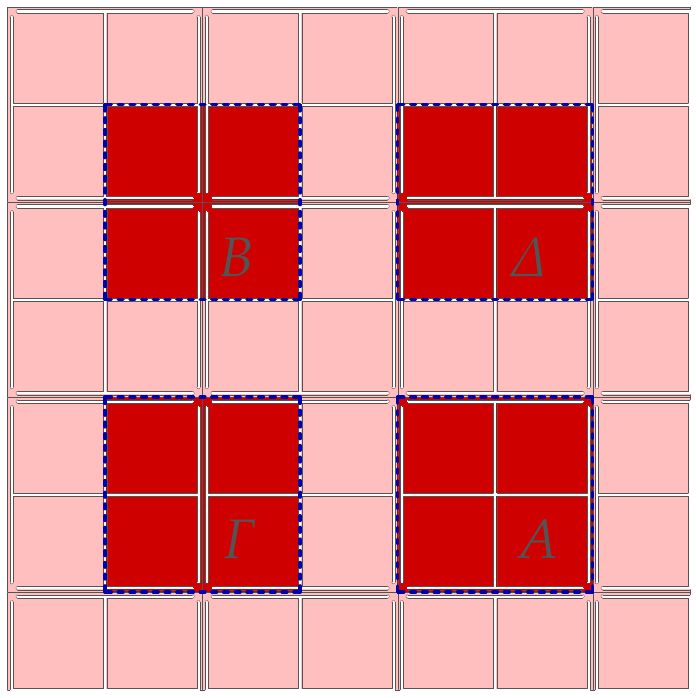}
	\includegraphics[width=0.4\textwidth]{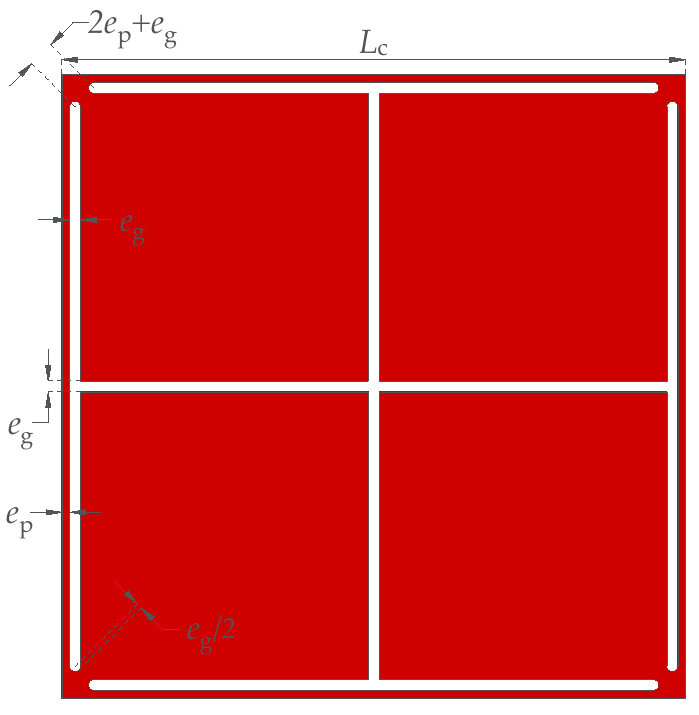}
	\vspace{0.3cm} 
	
	\begin{tabular}{|c|c|}
		\hline
		\textbf{Parameter} & \textbf{Value}\\
		\hline
		$L_{c}$ & 	\SI{50}{\milli\meter}  \\
		\hline
		$e_{p}$& $\SI{0.875}{\milli\meter}$ \\
		\hline 
		$e_{g}$ & \SI{0.625}{\milli\meter}\\
		\hline
	\end{tabular}
	\caption{(Left) the FR metamaterial with four choices of primitive unit cell ($A, B, \Gamma$ and $\Delta$) and (right and bottom) geometrical dimensions shown on unit cell $A$.}
	\label{fig:FR_unit_cell}
\end{figure}

Based on the Local Resonance mechanism, the Four-Resonator (FR) metamaterial was first designed in \cite{demore2022unfolding}.
This design also belongs to the plane crystallographic group $p4mm$.
The initially designed unit cell was unit cell $A$ in Fig.~\ref{fig:FR_unit_cell} while three more possible choices of unit cell are shown in the same figure with dark red and inside blue dashed-line squares.
The four resonators activate a Local Resonance mechanism which is responsible for the band-gap at low frequencies, as can be seen in the corresponding band structure in Fig.~\ref{fig:FR_unit_cell_DCs}.
For a dispersion plot including frequencies up to the end of the acoustic range see Appendix \ref{sec:band_str_FR_app}.
\begin{figure}[h]
	\centering
	\includegraphics[width=0.48\textwidth]{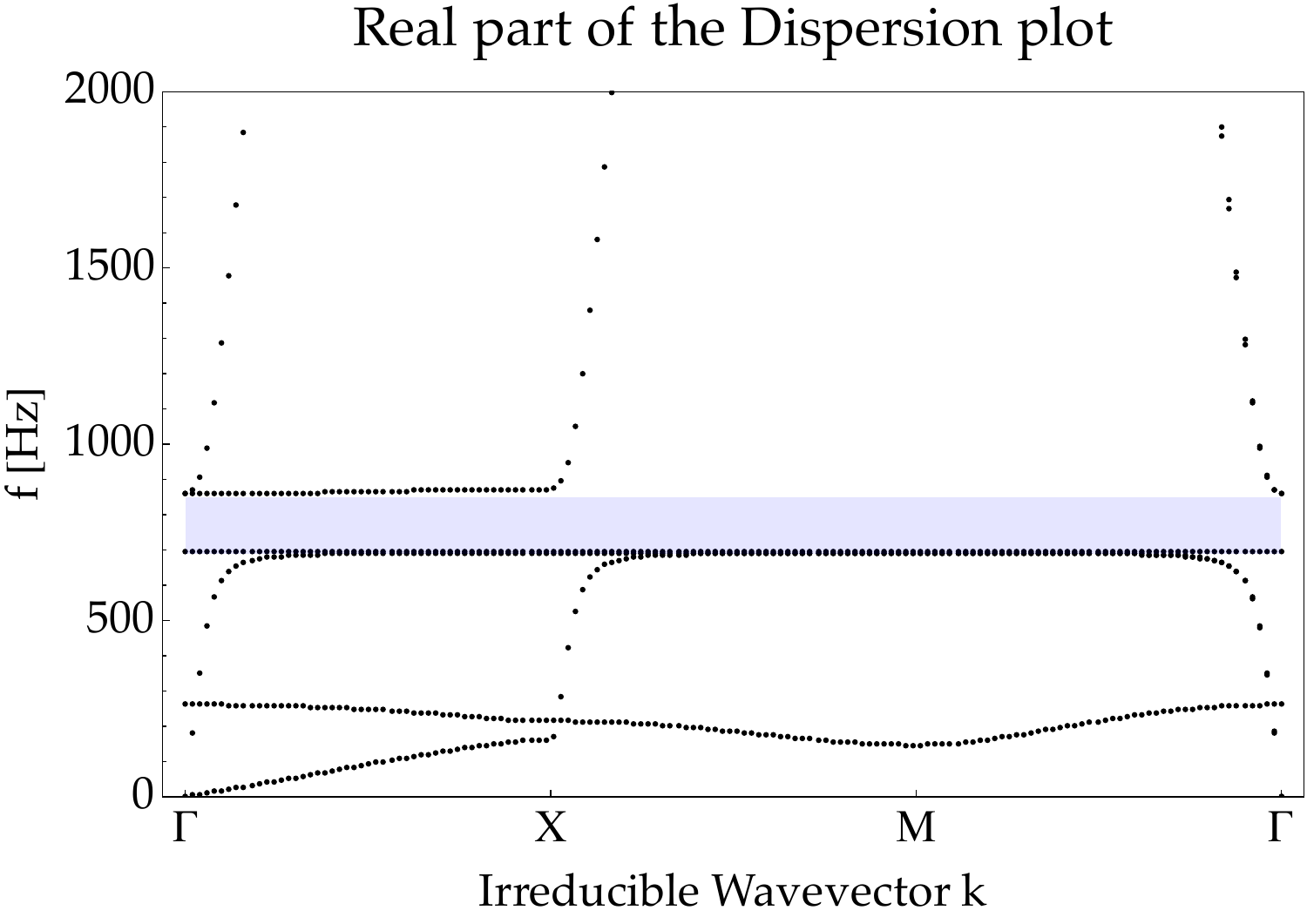}
	\includegraphics[width=0.48\textwidth]{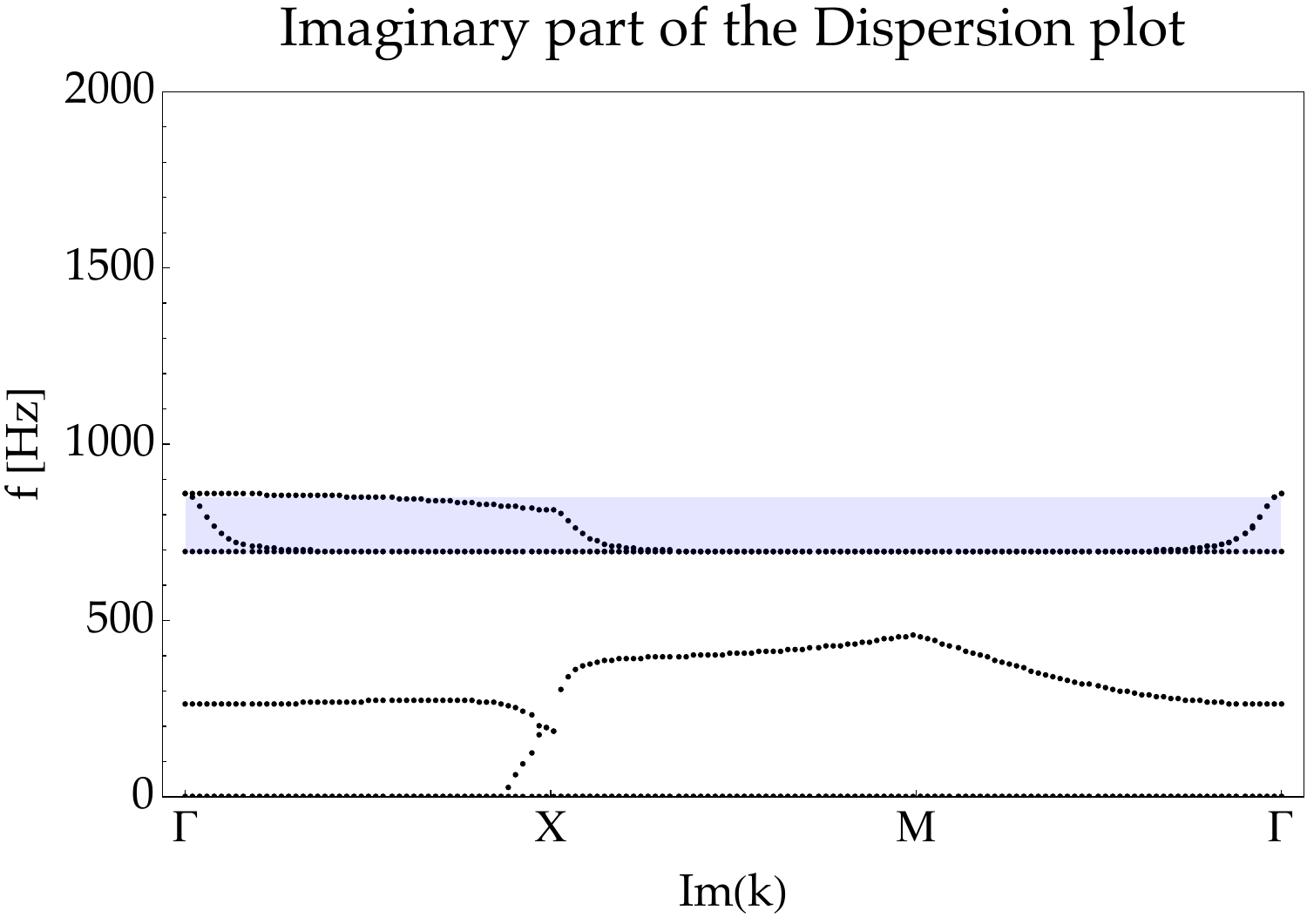}
	\caption{Band structure for the FR metamaterial with (left) real part and (right) imaginary part. The Local Resonance omnidirectional band-gap is highlighted in blue.}
	\label{fig:FR_unit_cell_DCs}
\end{figure}
%
\section{The Dual Cell method}\label{sec:dual_cell_method}
In this section we describe the dual cell method and apply it to the infinite chains of section \ref{sec:unit_cells}.
The method can be divided into two simple steps: \textit{(i)} choose two unit cells from the same parent metamaterial and \textit{(ii)}  couple the two unit cells in a specific way, creating a new ``dual'' unit cell which will be the unit cell of the new metamaterial.
The goal is to show that such a coupling can provide enhanced attenuation capabilities compared to the parent metamaterial, i.e. exhibit additional and in some cases more efficient band-gaps (in terms of attenuation) than those of the parent metamaterial.
In this section, the method is described using 1D spring-mass chains and it is then applied to 2D metamaterials in the next section. 

\begin{figure}[htbp]
	\centering
	
	\begin{tikzpicture}[
		spring/.style={
			thick,
			decorate,
			decoration={zigzag,segment length=6,amplitude=3}
		},
		mass1/.style={circle,draw,fill=blue!20,minimum size=8mm},
		mass2/.style={circle,draw,fill=red!20,minimum size=8mm}
		]
		
		\node at (-1,0) {$\cdots$};
		
		\node[mass2] (m1) at (0,0) {$m_2$};
		\node[mass1] (m2) at (2,0) {$m_1$};
		\node[mass1] (m3) at (4,0) {$m_1$};
		\node[mass2] (m4) at (6,0) {$m_2$};
		
		\node[mass2] (m5) at (8,0) {$m_2$};
		\node[mass1] (m6) at (10,0) {$m_1$};
		\node[mass1] (m7) at (12,0) {$m_1$};
		\node[mass2] (m8) at (14,0) {$m_2$};
		
		\draw[spring] (m1)-- node[above]{$k_1$} (m2);
		\draw[spring] (m2)-- node[above]{$k_1$} (m3);
		\draw[spring] (m3)-- node[above]{$k_1$} (m4);
		\draw[spring] (m4)-- node[above]{$k_1$} (m5);
		\draw[spring] (m5)-- node[above]{$k_1$} (m6);
		\draw[spring] (m6)-- node[above]{$k_1$} (m7);
		\draw[spring] (m7)-- node[above]{$k_1$} (m8);
		
		\node at (15.5,0) {$\cdots$};
		
		\node[blue] at (7,-1.5) {\small Dual Unit cell };
		
		\draw[blue, line width=1pt, dashed] (3,-1) rectangle (11,1);
		
	\end{tikzpicture}
	\vspace{1cm} 
	
	
	\begin{subfigure}[b]{0.48\textwidth}
		\centering
		\includegraphics[width=\textwidth]{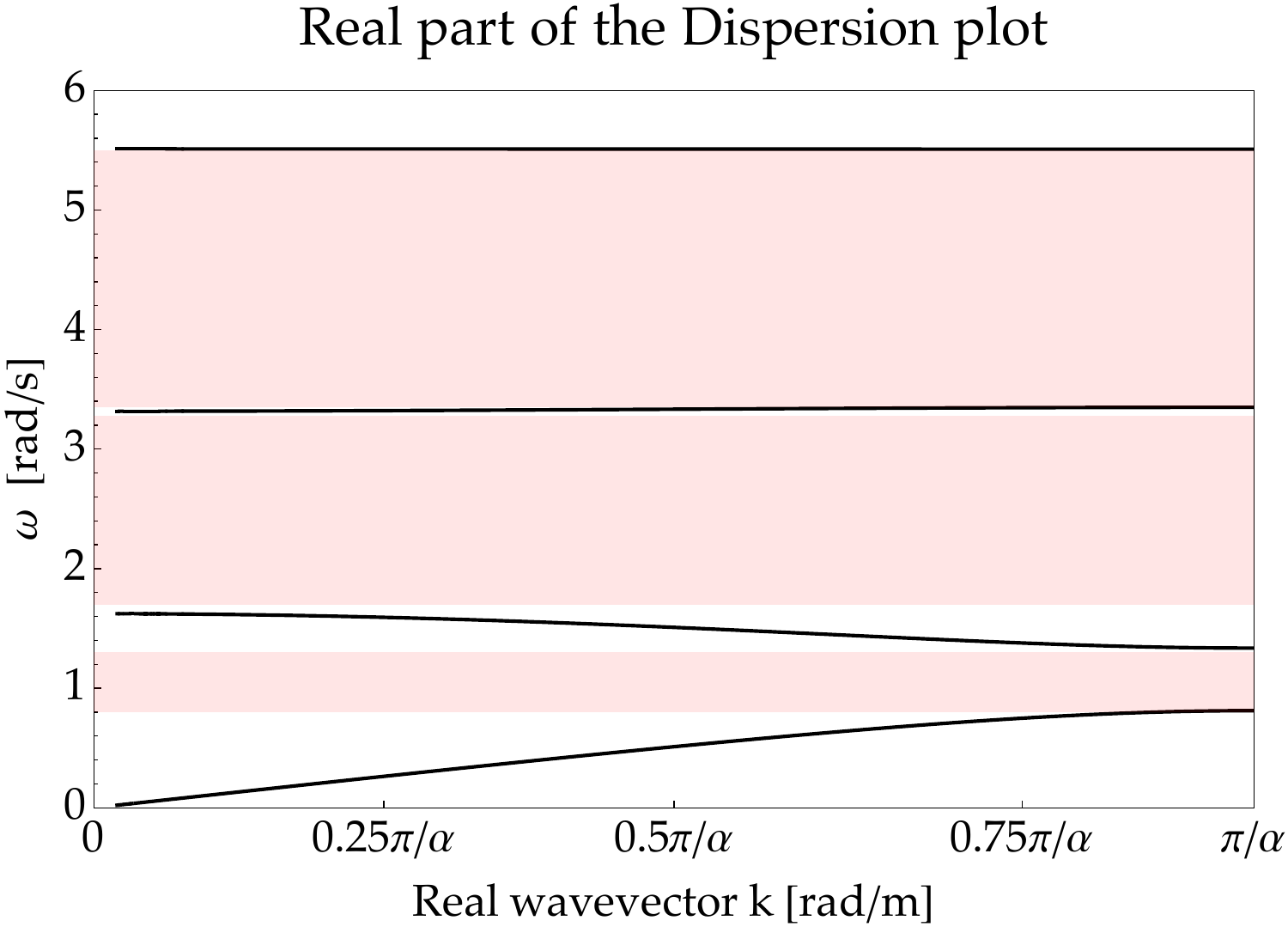}
	\end{subfigure}
	\hfill 
	\begin{subfigure}[b]{0.48\textwidth}
		\centering
		\includegraphics[width=\textwidth]{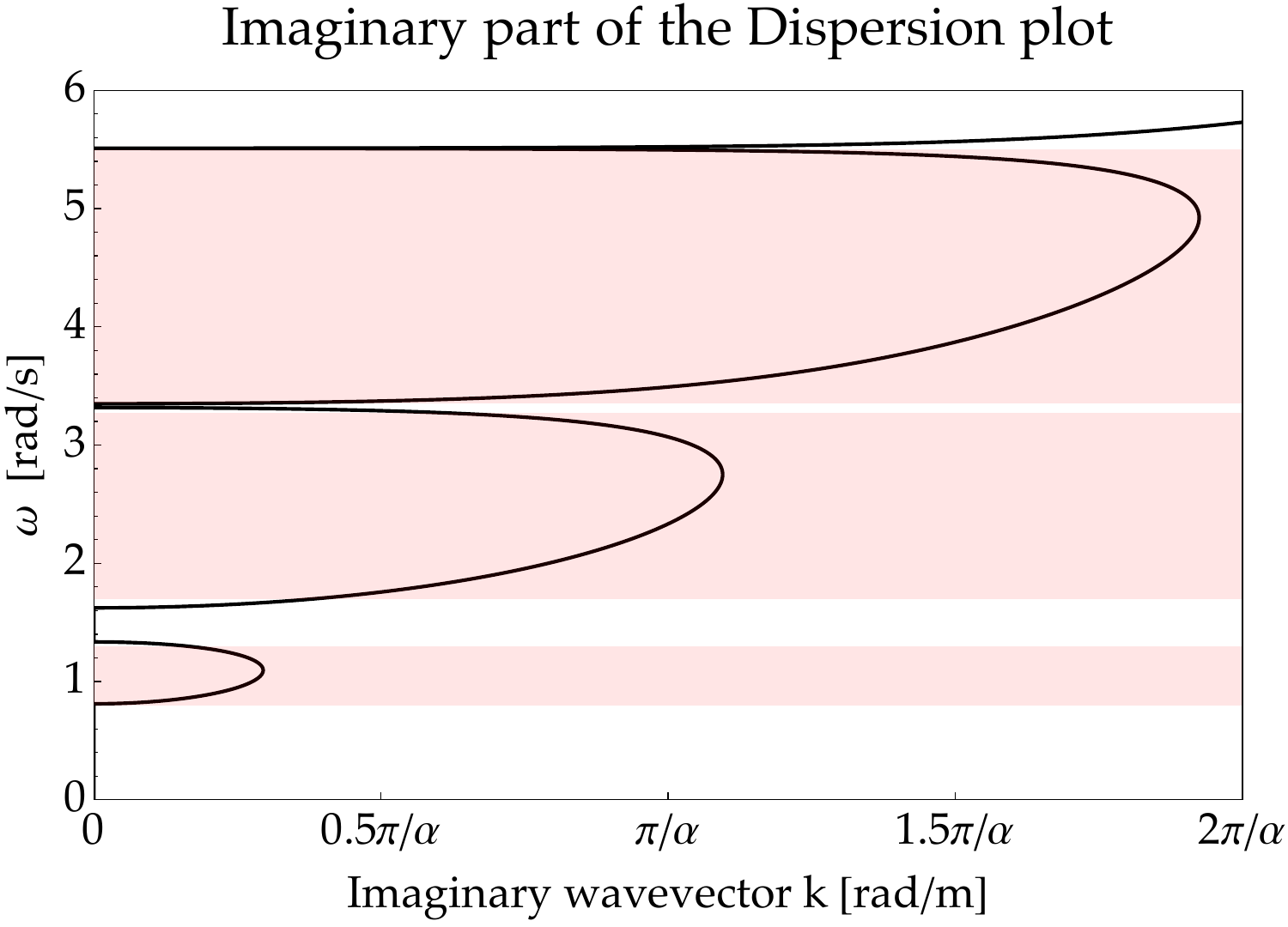}
	\end{subfigure}
	\caption{(Top) infinite tetra-atomic chain that results from applying the dual cell method on the diatomic chain shown in Fig.~\ref{fig:diatomic}, with the new dual unit cell shown in the dashed line box. (Bottom) real and imaginary parts of the dispersion plot with band-gaps highlighted in red.}
	\label{fig:tetraatomic}
	
\end{figure}

We start by applying the method on the diatomic chain of Fig.~\ref{fig:diatomic} which, as already mentioned, represents Bragg Scattering and can therefore be associated with the SCH metamaterial.
We simply couple the two unit cells shown in Fig.~\ref{fig:diatomic} (bottom) into a new dual unit cell, periodically repeat this dual cell in space and create a new chain, as shown in Fig.~\ref{fig:tetraatomic}.
This new tetra-atomic chain has a lattice constant twice the original, two extra degrees of freedom and it can exhibit two additional Bragg-type band-gaps with respect to the diatomic chain which also show enhanced attenuation, as can also be seen in Fig.~\ref{fig:tetraatomic} (notice how in the imaginary part of the plot we need to use a wavenumber with twice the value of the biggest wavevector to see the attenuation value of the highest Bragg gap).
For the parameter values and equilibrium equations in matrix form see Appendix \ref{sec:eq_eq}.
%
%

\begin{figure}[htbp]
	\centering
	\resizebox{0.8\textwidth}{!}{ 
		\begin{tikzpicture}[
			spring/.style={
				thick,
				decorate,
				decoration={zigzag, segment length=9, amplitude=4.5}
			},
			outerMass/.style={draw=black, line width=2.5pt},
			innerMass/.style={fill=red!20, draw=black, line width=1pt},
			cellBox/.style={draw=blue!70, dashed, line width=1pt, rounded corners=2pt}
			]
			
			\def\radiusOuterA{1.05}   
			\def\radiusOuterB{0.525}  
			\def\radiusInnerA{0.33}   
			\def\radiusInnerB{0.165}  
			
			\def\xA{0}
			\def\xB{2.475}
			\def\xC{5.325}
			\def\xD{7.8}
			
			\node[scale=1.5] at (-2.2, 0) {\Large $\dots$};
			\draw[spring] (-1.8, 0) -- (-\radiusOuterA, 0);
			\node[above] at ({(-1.8 + (-\radiusOuterA))/2}, 0.2) {$K_{1}$};
			
			\draw[outerMass] (\xA,0) circle (\radiusOuterA);
			\node[above] at (\xA, \radiusOuterA+0.2) {\large $M_{1}$};
			\draw[spring] (\xA-\radiusOuterA+0.01,0) -- (\xA-\radiusInnerA,0);
			\draw[innerMass] (\xA,0) circle (\radiusInnerA);
			\node[below] at (\xA, -\radiusInnerA-0.1) {$m_{1}$};
			\node[below] at (\xA-\radiusOuterA*0.6, -0.2) {$k_{1}$};
			
			\draw[spring] (\xA+\radiusOuterA,0) -- (\xB-\radiusOuterB,0);
			\node[above] at ({(\xA+\radiusOuterA+\xB-\radiusOuterB)/2}, 0.2) {$K_{1}$};
			
			\draw[outerMass] (\xB,0) circle (\radiusOuterB);
			\node[above] at (\xB, \radiusOuterB+0.2) {\large $M_{2}$};
			\draw[spring] (\xB-\radiusOuterB+0.01,0) -- (\xB-\radiusInnerB,0);
			\draw[innerMass] (\xB,0) circle (\radiusInnerB);
			\node[below] at (\xB, -\radiusInnerB+0.7) {$m_{2}$};
			\node[below] at (\xB-\radiusOuterB*0.6 +0.25, -0.025) {$k_{2}$};
			
			\draw[spring] (\xB+\radiusOuterB,0) -- (\xC-\radiusOuterB,0);
			\node[above] at ({(\xB+\radiusOuterB+\xC-\radiusOuterB)/2}, 0.2) {$K_{2}$};
			
			\draw[outerMass] (\xC,0) circle (\radiusOuterB);
			\node[above] at (\xC, \radiusOuterB+0.2) {\large $M_{2}$};
			\draw[spring] (\xC-\radiusOuterB+0.01,0) -- (\xC-\radiusInnerB,0);
			\draw[innerMass] (\xC,0) circle (\radiusInnerB);
			\node[below] at (\xC, -\radiusInnerB+0.7) {$m_{2}$};
			\node[below] at (\xC-\radiusOuterB*0.6+0.25, -0.025) {$k_{2}$};
			
			\draw[spring] (\xC+\radiusOuterB,0) -- (\xD-\radiusOuterA,0);
			\node[above] at ({(\xC+\radiusOuterB+\xD-\radiusOuterA)/2}, 0.2) {$K_{1}$};
			
			\draw[outerMass] (\xD,0) circle (\radiusOuterA);
			\node[above] at (\xD, \radiusOuterA+0.2) {\large $M_{1}$};
			\draw[spring] (\xD-\radiusOuterA+0.01,0) -- (\xD-\radiusInnerA,0);
			\draw[innerMass] (\xD,0) circle (\radiusInnerA);
			\node[below] at (\xD, -\radiusInnerA-0.1) {$m_{1}$};
			\node[below] at (\xD-\radiusOuterA*0.6, -0.2) {$k_{1}$};
			
			\draw[spring] (\xD+\radiusOuterA,0) -- (\xD+\radiusOuterA+1.2,0);
			\node[above] at (\xD+\radiusOuterA+0.6, 0.2) {$K_{1}$};
			\node[scale=1.5] at (\xD+\radiusOuterA+1.7, 0) {\Large $\dots$};
			
			\draw[cellBox]
			({(\xA+\xB)/2 - 2.7}, -1.3)
			rectangle
			({(\xC+\xD)/2 - 0.2},  1.6);
			\node[blue!70, below] at ({((\xA+\xB)/2 + (\xC+\xD)/2)/2}, -1.3)
			{Dual Unit Cell};
			
		\end{tikzpicture}
	}
	\vspace{1cm} 
	
	
	\begin{subfigure}[b]{0.48\textwidth}
		\centering
		\includegraphics[width=\textwidth]{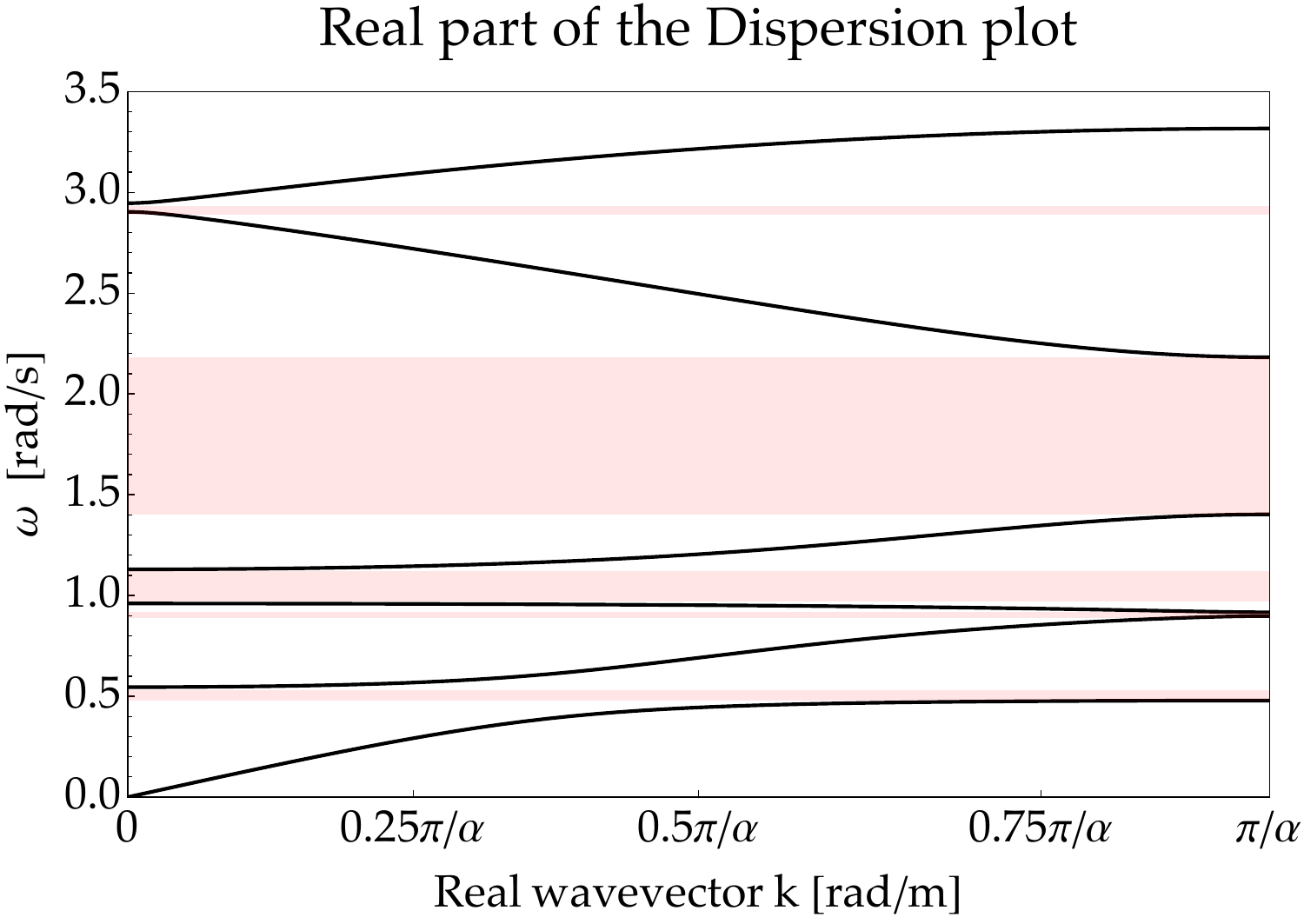}
	\end{subfigure}
	\hfill 
	\begin{subfigure}[b]{0.48\textwidth}
		\centering
		\includegraphics[width=\textwidth]{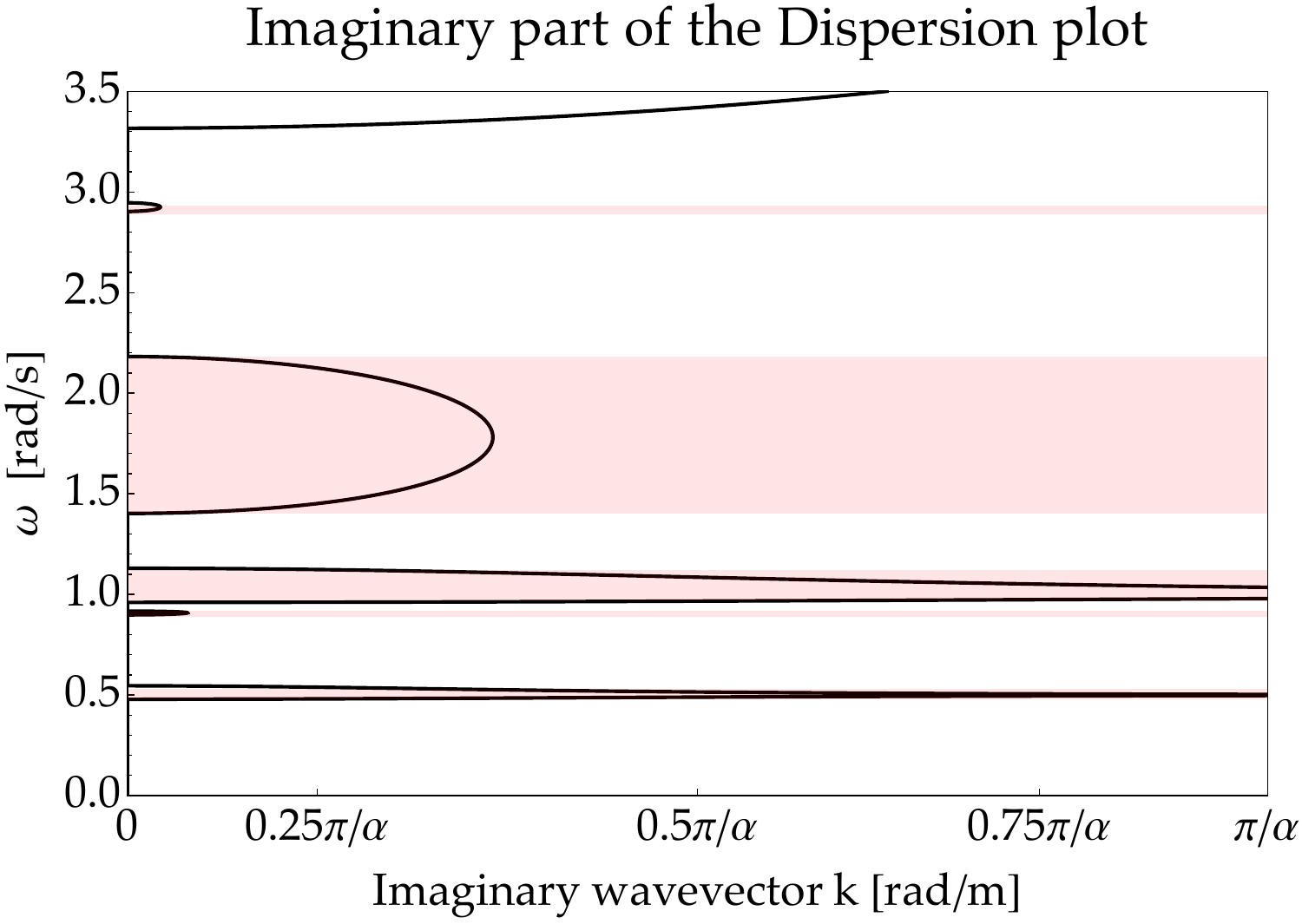}
	\end{subfigure}
	\caption{ (Top) infinite triatomic mass-in-mass chain that results from applying the dual cell method on the simple mass-in-mass chain of Fig.~\ref{fig:mass_in_mass}, with the new dual unit cell shown in the dashed line box. (Bottom) real and imaginary parts of the dispersion plot with band-gaps highlighted in red.}
	\label{fig:dual_from_mass_in_mass}
\end{figure}
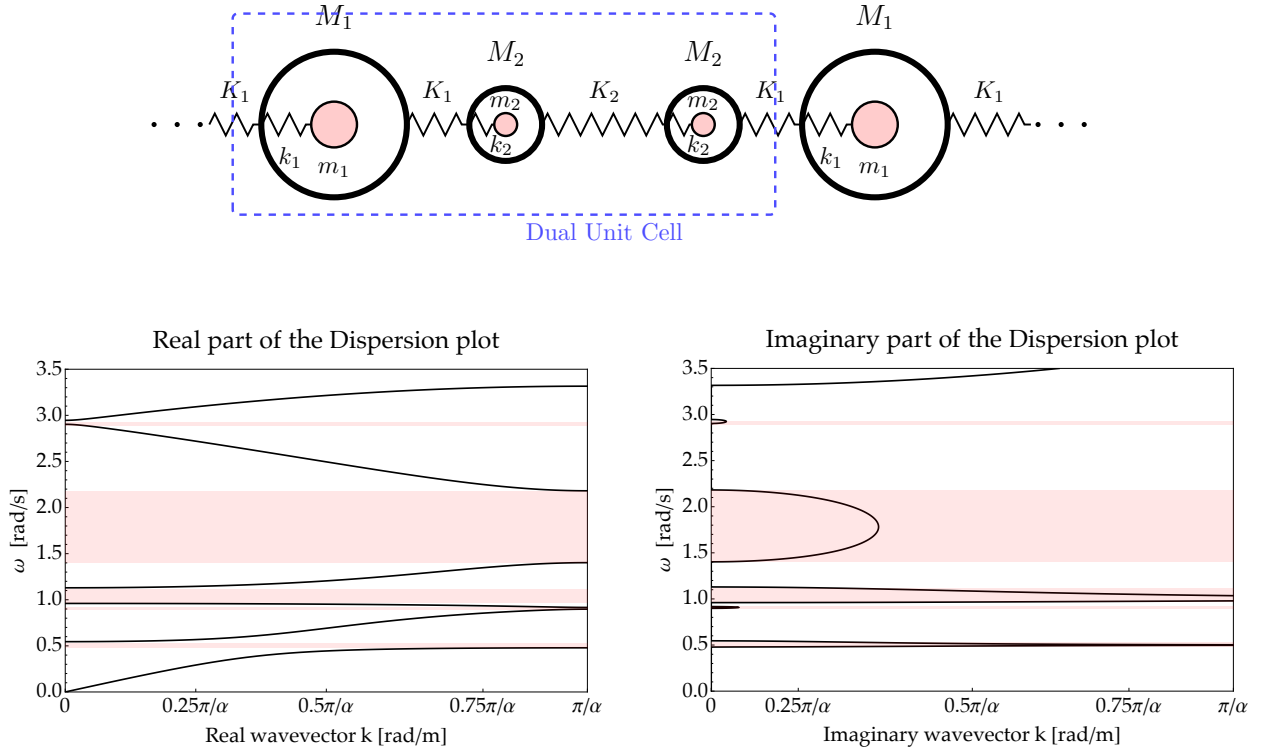

The same method can be applied to the mass-in-mass chain  of Fig.~\ref{fig:mass_in_mass}.
The coupling of the two unit cells shown in Fig.~\ref{fig:mass_in_mass} creates a new dual unit cell whose periodic repetition creates a new chain shown in Fig~\ref{fig:dual_from_mass_in_mass}.
Notice how unit cell $B$ in Fig.~\ref{fig:mass_in_mass} has springs with half the number of coilings, which means half the total spring length and twice the total stiffness, while unit cell $A$ has halved masses (both primary and resonator masses) and halved resonator springs\footnote{Given that this is a 1D model with a 2D sketch and the resonator springs could be placed in different configurations, what happens to the resonator springs during the unit cell choice procedure is a result of this sketch.
	Using symmetry considerations, for both resonator springs we choose a stiffness value $k$ twice the initial, such that the spring length was halvened.
	We could have arguably chose not to modify their stiffness value or even disregard one resonator spring together with the mass it carries, which would make the dispersion behavior less rich of band-gaps, but still the dual cell method would produce a better dispersion plot than the initial mass-in-mass chain, in the sense that it would produce additional band-gaps.}.
The new chain shown in Fig~\ref{fig:dual_from_mass_in_mass} is a hybrid triatomic-locally resonant chain with six degrees of freedom and the possibility for four more band-gaps, three of which, in this specific case, are Bragg-type gaps and one of Local Resonance type.
In this case the additional Local Resonance gap shows enhanced attenuation while the Bragg gaps do not show particularly enhanced attenuation values.
For the parameter values and equations of motion in matrix form see Appendix \ref{sec:eq_eq}.

We note that for the dual cell method applied on the diatomic and mass-in-mass chains in this section, unit cells that are only related by a half-period shift were used.
This particular value of shift is, however, only a choice and can be varied to end up with sometimes different infinite chains with different degrees of freedom.
This implies a different number of band-gaps and different attenuation profiles dictated solely by the value of this shift. 
Additionally it must be noted that in both cases shown in this section, the dual cell method produces a system with more band-gaps, some of which also exhibit enhanced attenuation profiles, with the cost being the enlargement of the lattice constant.
This has some interesting implications on the design of band-gap metamaterials that are discussed in section \ref{sec:concl}, together with a
way in which the enlargement of the lattice constant can sometimes be avoided.
\section{The Dual Cell method in 2D}\label{sec:dual_2D}

We now wish to apply the dual cell method in 2D and specifically to the parent metamaterials introduced in section~\ref{sec:unit_cells}.
With the addition of the second dimension, the coupling of unit cells can now be done in a number of different configurations, such that the resulting infinitely extended metamaterial can belong either to the same or to a different plane crystallographic group than the initial parent metamaterial.
It has been shown in \cite{maurin2018probability} that the higher the symmetry of the crystallographic group of the metamaterial, the higher the probability is, that the band-gap extrema are located on the Irreducible Brillouin Contour.
Given the complexity of the microstructures that results from these coupling configurations, we chose to use parent metamaterials of the $p4mm$ wallpaper group (SCH and FR metamaterials), and couple the unit cells coming from these materials in such a way, so that high symmetry remains also in the metamaterials that result from the dual cell method.
This simplifies the band structure calculations saving computational time and at the same time confines the study to highly symmetric designs that have been more often associated to omnidirectional ultra-wide band-gaps in several cases \cite{yuksel2015shape,nadejde2023pushing, xiong2024bandgap}.

Two different coupling configurations are examined.
We first start with the most intuitive way, using a side-by-side coupling which produces a $p2mm$ wallpaper group and we show that this configuration produces multiple band-gaps in only one propagation direction, with an enhanced attenuation profile (see section~\ref{sec:p2mm}).
The second configuration is that of a chessboard pattern, with a limited choice of unit cells such that the resulting metamaterial belongs to the $p4mm$ wallpaper group.
This configuration is found to be superior to the side-by-side one, since it produces more omnidirecitonal band-gaps (see section~\ref{sec:chess}).

Since the base material for both parent metamaterials is structural steel, all band structure calculations presented in this section have been performed assuming linearly elastic material behavior and that the plane strain assumption holds. 
More specifically, the Structural Mechanics Module of \comsol was used and a mesh convergence study was carried out in all cases.

\subsection{Side-by-side configuration}\label{sec:p2mm}
In this section, the side-by-side coupling/configuration of unit cells is examined.
This particular side-by-side coupling creates a motif with a rectangular Bravais lattice.
According to the findings in \cite{maurin2018probability}, from the wallpaper groups that correspond to a rectangular lattice, the one found to have the highest probability of band-gap extrema located on the IBC is the $p2mm$.
As such, we proceed with choices of unit cell that create a dual cell whose periodic repetition in the plane creates a metamaterial of the $p2mm$ wallpaper group (see Figs.~\ref{fig:SCH_p2mm_UC_DCs} and~\ref{fig:FR_p2mm_UC_DCs}).

Given the four choices of unit cells in Figs.~\ref{fig:SCH_unit_cell} and~\ref{fig:FR_unit_cell} ($A, B, \Gamma$ and $\Delta$) the coupling of unit cells can theoretically be done in six different ways ($AB, A\Gamma, A\Delta, B\Gamma, B\Delta $ and $\Gamma\Delta$).
However, the dual cell $B\Gamma$ is geometrically incompatible for a side-by-side configuration, while the cell $AB$ is a shifted version of $\Gamma\Delta$ which means they produce the same band structure in a Bloch-Floquet Analysis. 
The same holds for the cells $A\Gamma$ and $B\Delta$: one is a shifted version of the other.
This leaves us with only three unique dual cells: $AB, A\Gamma, $ and $A\Delta$.
The dual cell that produces the best band structure for both the $p2mm$ SCH and FR metamaterials is the $AB$ which is shown in this Section, while the band structure for the rest can be seen in Appendix~\ref{sec:band_str_p2mm}.
\subsubsection{Side-by-side configuration for the SCH metamaterial}\label{sec:dual_SCH}
\begin{figure}[h!]
	\centering
	\includegraphics[width=0.55\textwidth]{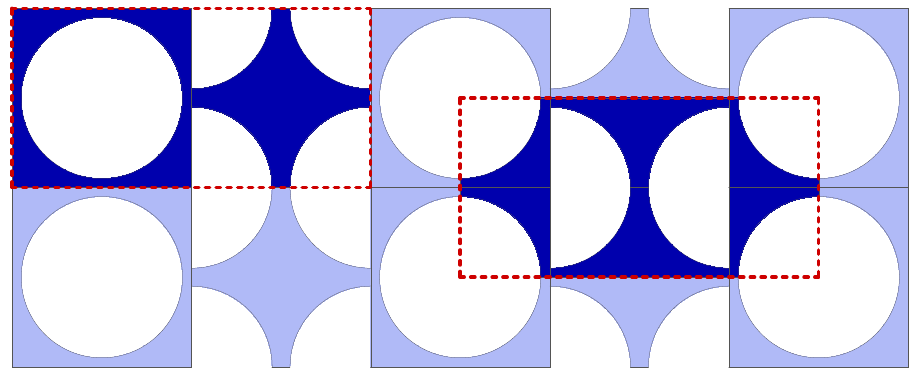}
	\includegraphics[width=0.4725\textwidth]{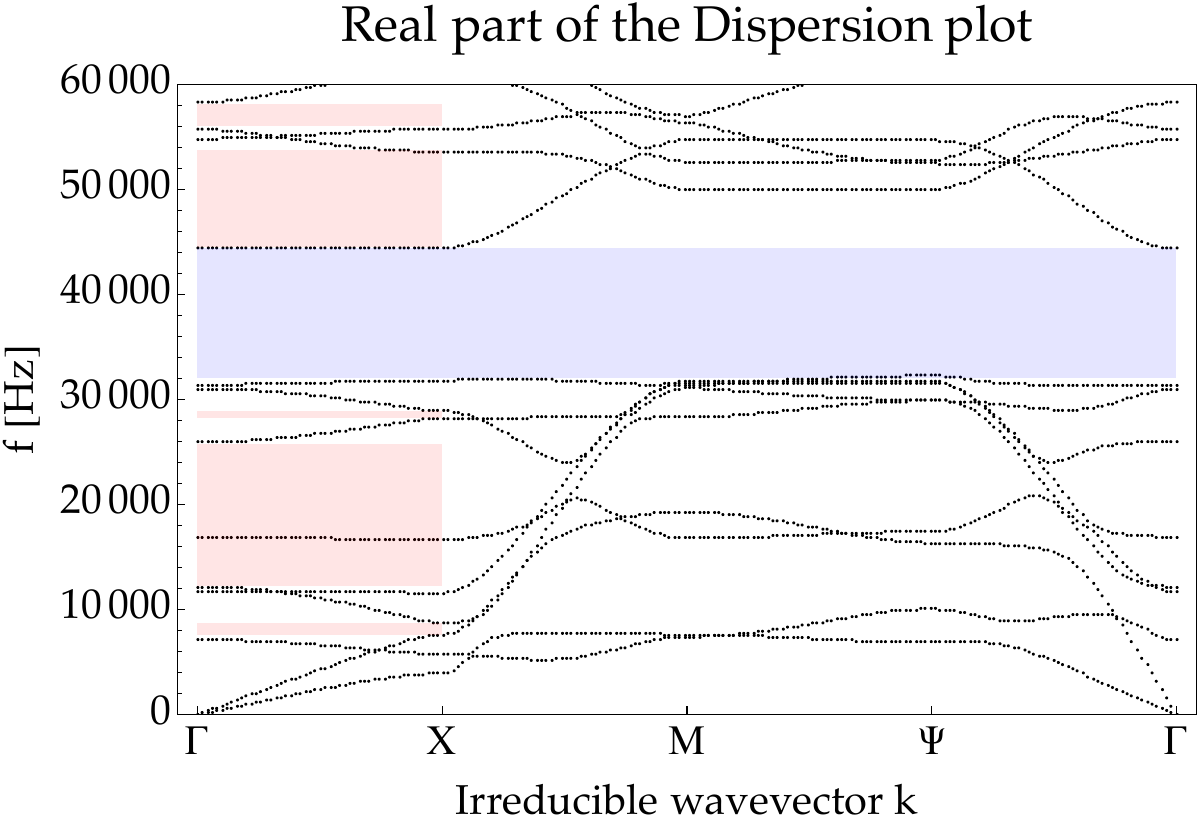}
	\includegraphics[width=0.4875\textwidth]{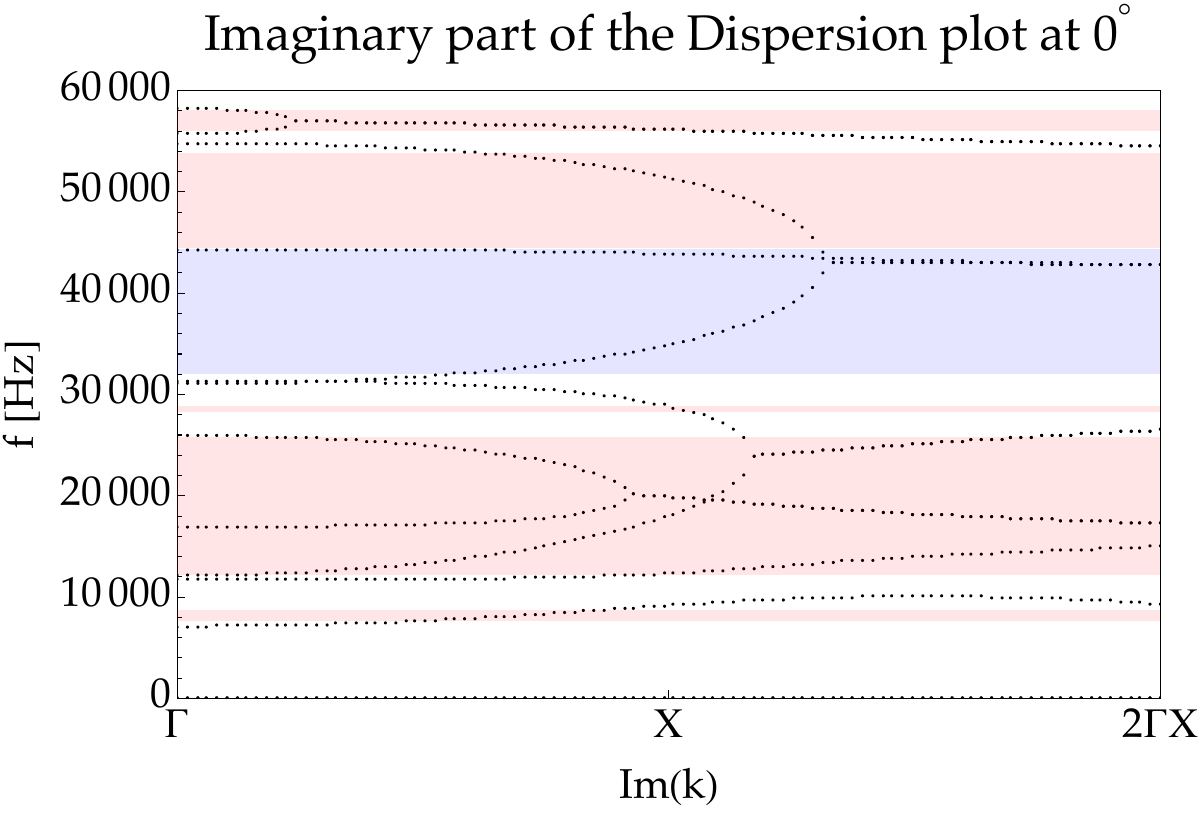}
	\caption{(Top) the SCH $p2mm$ design that results by applying the dual cell method on unit cells $A$ and $B$  of Fig.~\ref{fig:SCH_unit_cell} using a side-by-side coupling. Two primitive unit cells are highlighted with dark blue. The first (left) shows the $AB$ coupling in a clear way while the second (right) carries the full symmetry of the $p2mm$ wallpaper group. (Bottom) corresponding band structure with (left) real part and (right) imaginary part only for wave propagation along $0^{\circ} $. In the imaginary part of the plot the wavevector used reaches up to two times the irreducible wavevector's maximum size. Omnidirectional band-gaps are highlighted in blue while directional band-gaps at $0^{\circ} $ are highlighted in red.}
	\label{fig:SCH_p2mm_UC_DCs}
\end{figure}

In Fig.~\ref{fig:SCH_p2mm_UC_DCs} the SCH $p2mm$ design can be seen, which results by coupling the $A$ and $B$ unit cells of the SCH metamaterial using a side-by-side configuration. On the left, a primitve unit cell that shows the configuration in a clear way is highlighted in dark blue inside a red dashed-line square, while on the right another primitive unit cell is highlighted in the same way, which carries the full symmetry of the $p2mm$ wallpaper group. The corrsesponding band structure can be seen on the bottom.
%
\subsubsection{Side-by-side configuration for the FR metamaterial}\label{sec:dual_FR}
\begin{figure}[!htbp]
	\centering
	\includegraphics[width=0.55\textwidth]{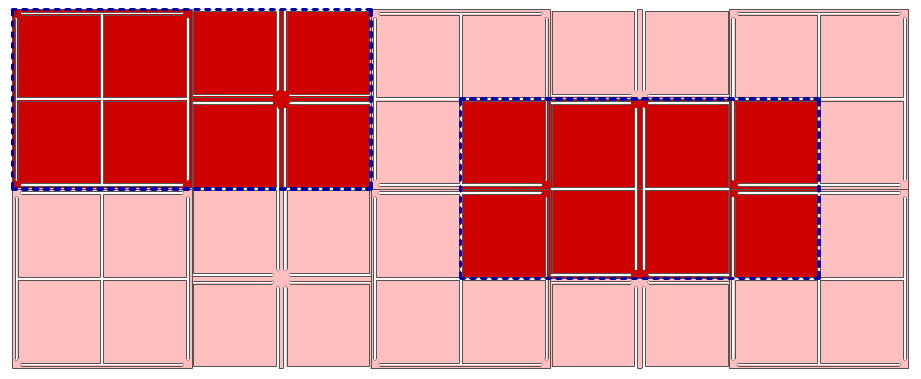}
	\includegraphics[width=0.4725\textwidth]{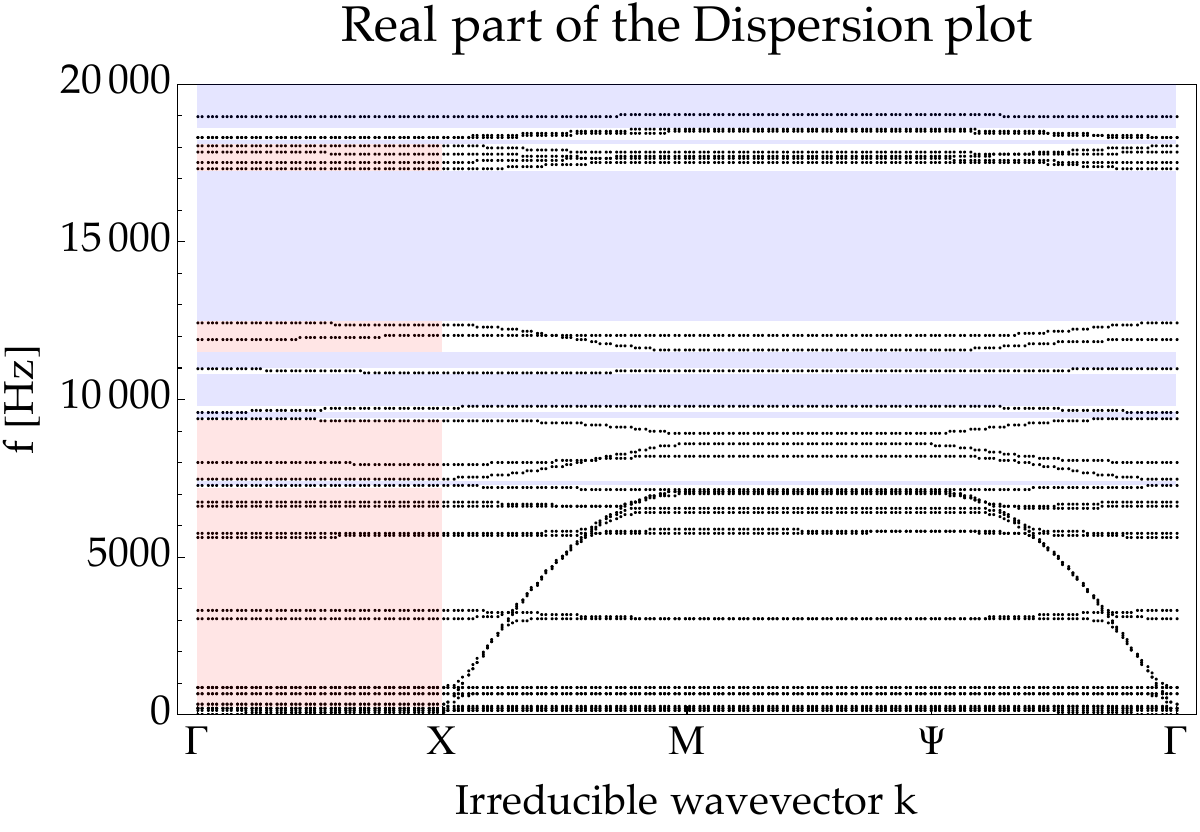}
	\includegraphics[width=0.4875\textwidth]{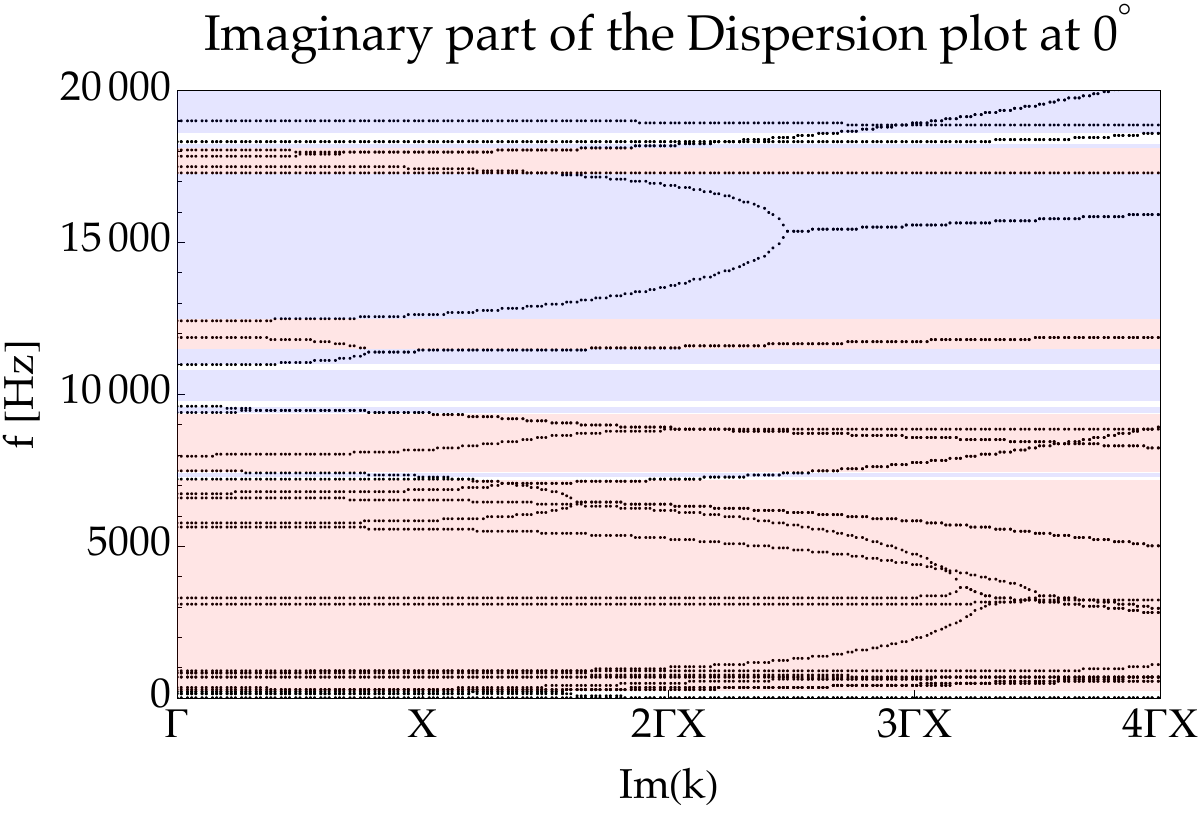}
	\caption{(Top) the FR $p2mm$ design that results by applying the dual cell method on unit cells $A$ and $B$  of Fig.~\ref{fig:FR_unit_cell} using a side-by-side coupling. Two primitive unit cells are highlighted with dark red. The first (left) shows the $AB$ coupling in a clear way while the second (right) carries the full symmetry of the $p2mm$ wallpaper group. (Bottom) corresponding band structure with (left) real part and (right) imaginary part only for wave propagation along $0^{\circ} $.  In the imaginary part of the plot the wavevector used reaches up to four times the irreducible wavevector's maximum size. Omnidirectional band-gaps are highlighted in blue while directional band-gaps at $0^{\circ} $ are highlighted in red.}
	\label{fig:FR_p2mm_UC_DCs}
\end{figure}

\begin{figure}[!htbp]
	\centering
	\includegraphics[width=0.4725\textwidth]{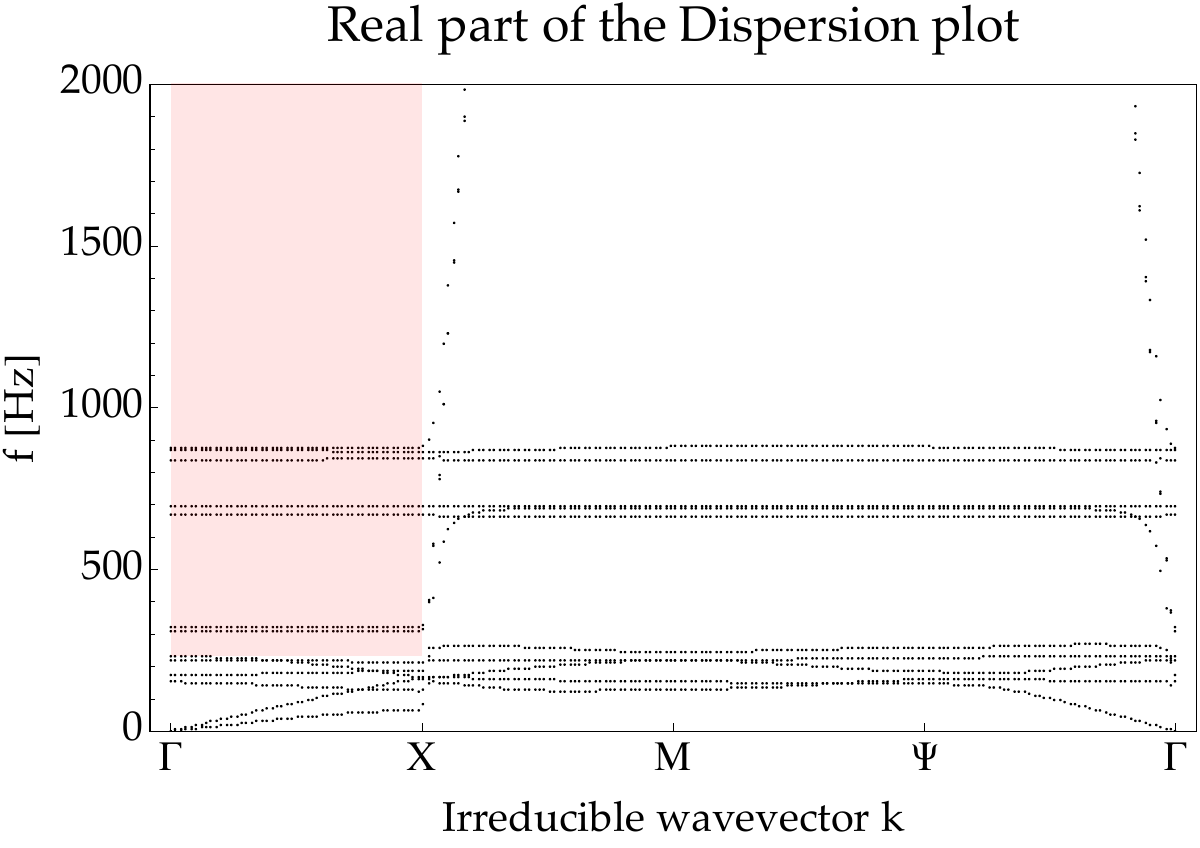}
	\includegraphics[width=0.4875\textwidth]{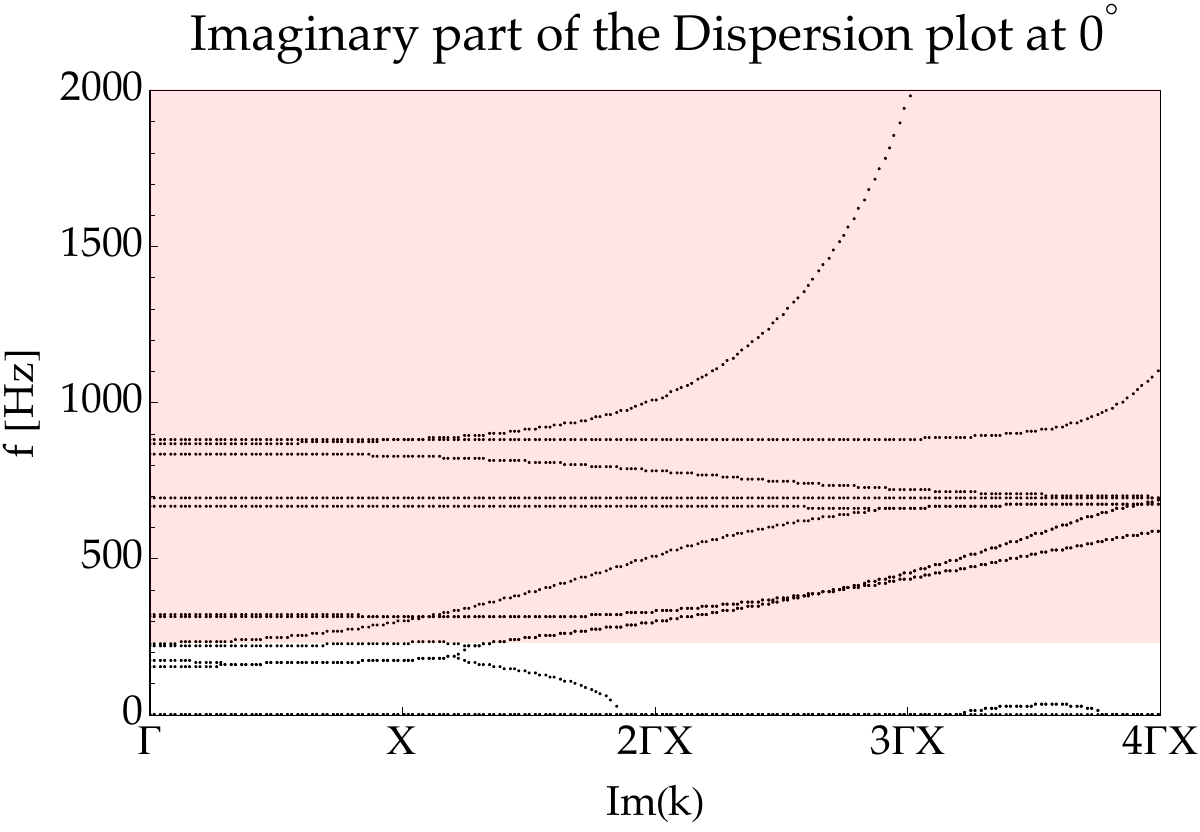}
	\caption{Band Structure for the FR $p2mm$ design of Fig.~\ref{fig:FR_p2mm_UC_DCs} highlighting low frequencies with (left) real part and (right) imaginary part only for wave propagation along $0^{\circ} $. In the imaginary part of the plot the wavevector used reaches up to four times the irreducible wavevector's maximum size. Directional band-gaps at  $0^{\circ} $ are highlighted in red.}
	\label{fig:FR_p2mm_DC_low}
\end{figure}

%
%
In Fig.~\ref{fig:FR_p2mm_UC_DCs} the FR $p2mm$ design can be seen, which results by coupling the $A$ and $B$ unit cells of the FR metamaterial using a side-by-side configuration. On the left, a primitve unit cell that shows the configuration in a clear way is highlighted in dark red inside a blue dashed-line square, while on the right another primitive unit cell is highlighted in the same way, which carries the full symmetry of the $p2mm$ wallpaper group. The corrsesponding band structure can be seen on the bottom.
An extra detail of the band structure for low frequencies can be seen in Fig.~\ref{fig:FR_p2mm_DC_low}.

\subsubsection{Discussion}\label{sec:side_by_side_disc}

In Figs.~\ref{fig:SCH_p2mm_UC_DCs}, ~\ref{fig:FR_p2mm_UC_DCs} and~\ref{fig:FR_p2mm_DC_low} it is observed that the side-by-side $AB$ configuration exhibits multiple band-gaps with enhanced attenuation for both parent metamaterials examined starting from low frequencies (notice how in the imaginary part of the dispersion plots at $0^{\circ} $ we need to use wavevectors with size bigger than the maximum size of the irreducible wavevector in order to see the actual attenuation values for some frequency ranges).
These band-gaps appear mostly in the  $0^{\circ} $ propagation direction, while only few omnidirectional band-gaps appear which are at high frequencies.
For the SCH $AB$ $p2mm$ design all band-gaps are Bragg-type gaps as can be seen in the corresponding imaginary part of the dispersion plot in Fig.~\ref{fig:SCH_p2mm_UC_DCs}.
For the FR $AB$ $p2mm$ design most of the band-gaps are again Bragg type as can be seen in Fig.~\ref{fig:FR_p2mm_UC_DCs}, but Local Resonance band-gaps also appear at low frequencies since the parent metamaterial is of the Local Resonance type (see the frequency range 250-900 Hz in Fig.~\ref{fig:FR_p2mm_DC_low}).
These results are of interest mainly for attenuating waves in the $0^{\circ}$ propagation direction. As our main goal is the opening of omnidirectional band-gaps so that waves can be attenuated in any direction, an alternative coupling configuration is needed.
\subsection{Chessboard configuration}\label{sec:chess}
With the aim of turning directional band-gaps of previous section to omnidirectional, a chessboard configuration is examined in this section.
In order to keep a high symmetry in the resulting metamaterial, we only couple unit cells that carry the highest symmetry of the parent metamaterial (the one of the $p4mm$ wallpaper group), since a chessboard pattern of two unit cells that carry the full symmetry of the $p4mm$ wallpaper group also results in a wallpaper group $p4mm$.
For any $p4mm$ metamaterial, there are only two primitive unit cells with a square shape that carry its full symmetry (in our case these are cells $A$ and $B$ in Figs.~\ref{fig:SCH_unit_cell} and~\ref{fig:FR_unit_cell})\footnote{Unit cells  $\Gamma$ and $\Delta$ do not carry the full symmetry of the wallpaper group and including any of them in a chessboard configuration would turn the highest rotational symmetry of the wallpaper group from 4-fold  to 2-fold, decreasing the overall symmetry of the chessboard pattern's wallpaper group.}.
This implies that there can only be one version of a $p4mm$ chessboard, where $A$ and $B$ act as the white and black squares of the chessboard respectively.

The chessboard pattern for both the SCH and FR metamaterials creates vertex connections at the points where the corners of the white squares of the chessboard would meet, since unit cell $A$ has a continuous outside boundary.
In order to avoid connecting geometry at a single point, we slightly modify unit cell $A$ by drilling small holes at the corners\footnote{These circular holes have a radius of \SI{0.25}{\milli\meter} and can be seen in both Figs.~\ref{fig:SCH_chess_UC_DCs} and~\ref{fig:FR_chess_UC_DCs} (notice how the corners of adjacent parent unit cells $A$ are disconnected and parent unit cells $B$ have a small hole in their center).},  such that both the symmetry and band structure of the parent metamaterial remain unaltered. 
Even if the drilling of holes does not alter the  band structure, the geometry of the rest choices of unit cell are of course modified for consistency, since they are unit cells of the same metamaterial.
\subsubsection{Chessboard configuration for the SCH metamaterial}\label{sec:SCH_chess}
\begin{figure}[!htbp]
	\centering
	\includegraphics[width=0.55\textwidth]{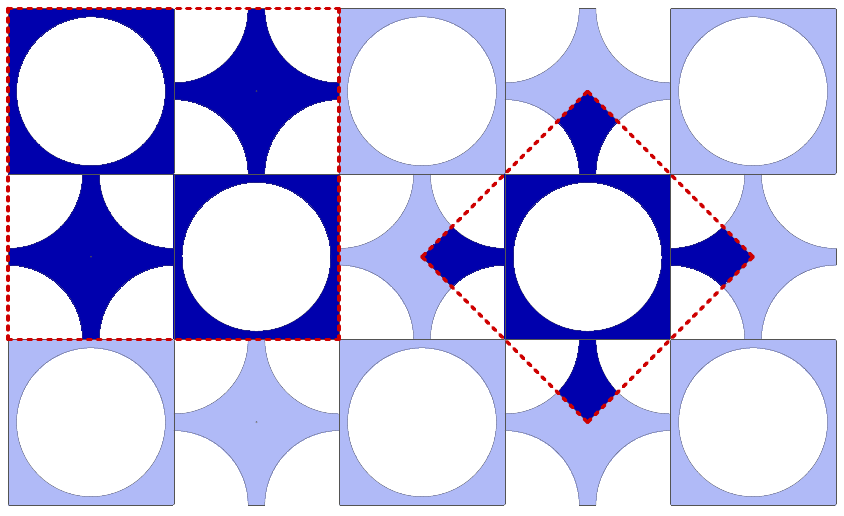}
	\includegraphics[width=0.475\textwidth]{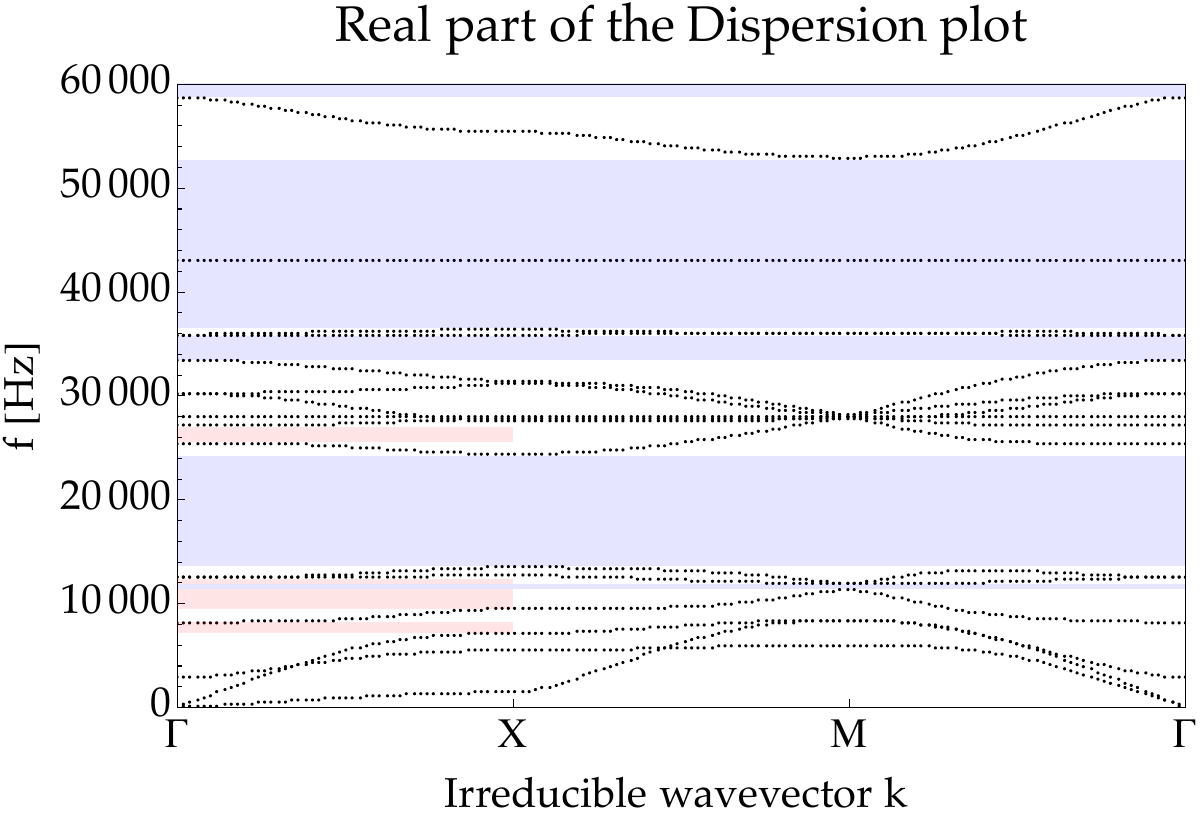}
	\includegraphics[width=0.4875\textwidth]{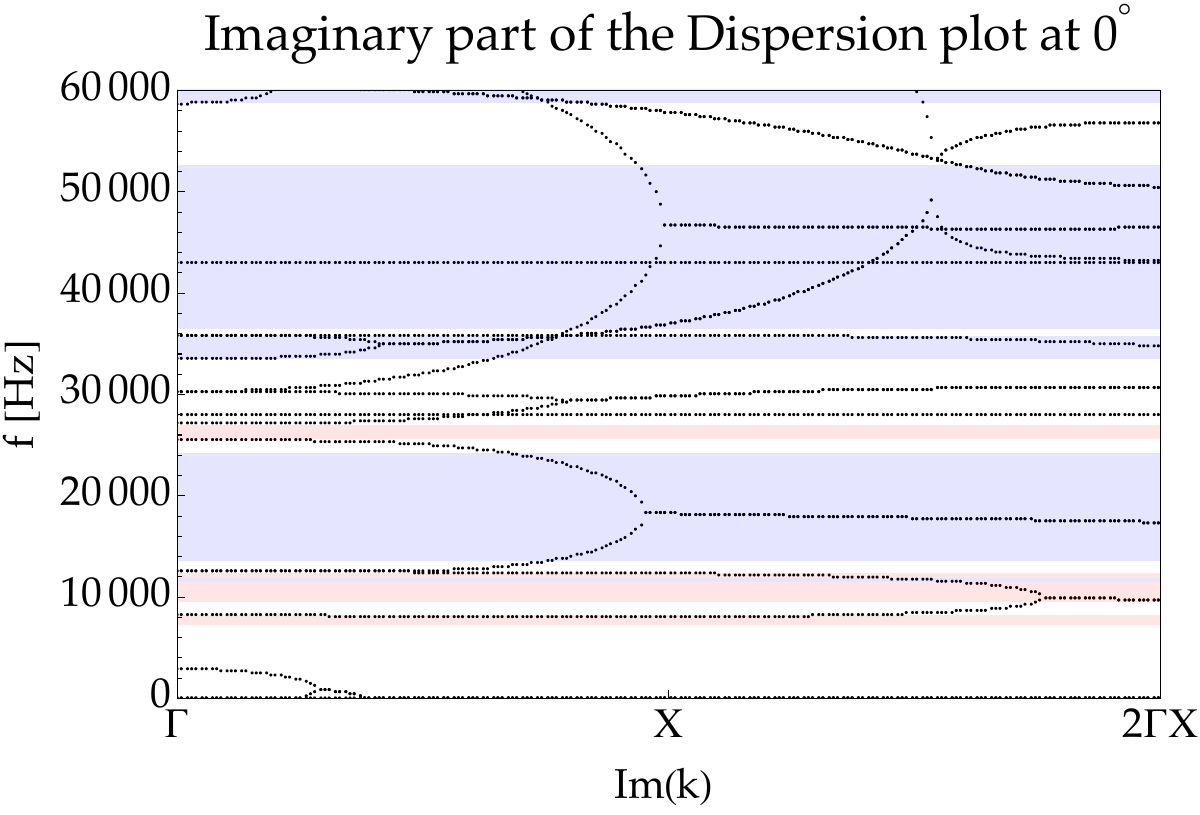}
	\caption{(Top) the SCH $p4mm$ design that results by applying the dual cell method on unit cells $A$ and $B$  of Fig.~\ref{fig:SCH_unit_cell} using a chessboard configuration. Two unit cells are highlighted with dark blue. The first (left) is a conventional unit cell that shows the $AB$ coupling in a clear way while the second (right) is a primitive unit cell that also carries the full symmetry of the $p4mm$ wallpaper group. (Bottom) corresponding band structure with (left) real part and (right) imaginary part for wave propagation along $0^{\circ} $. In the imaginary part of the plot the wavevector used reaches up to two times the irreducible wavevector's maximum size. Omnidirectional band-gaps are highlighted in blue while directional band-gaps at  $0^{\circ} $ are highlighted in red.}
	\label{fig:SCH_chess_UC_DCs}
\end{figure}

The chessboard SCH $p4mm$ metamaterial that results from applying the dual cell method through a chessboard configuration can be seen in Fig.~\ref{fig:SCH_chess_UC_DCs} (top) with two choices of unit cell highlighted with dark blue inside red dashed-line squares.
On the left, a conventional unit cell that shows the coupling in a clear way can be seen and on the right a primitive unit cell with the full symmetry of the $p4mm$ wallpaper group. 
The corresponding band structure can be seen at the bottom. 
Bloch-Floquet analysis that produces this band structure is done on the primitve unit cell shown, but rotated by $45^{\circ} $. 
\subsubsection{Chessboard configuration for the FR metamaterial}\label{sec:chess_FR}
\begin{figure}[h!]
	\centering
	\includegraphics[width=0.55\textwidth]{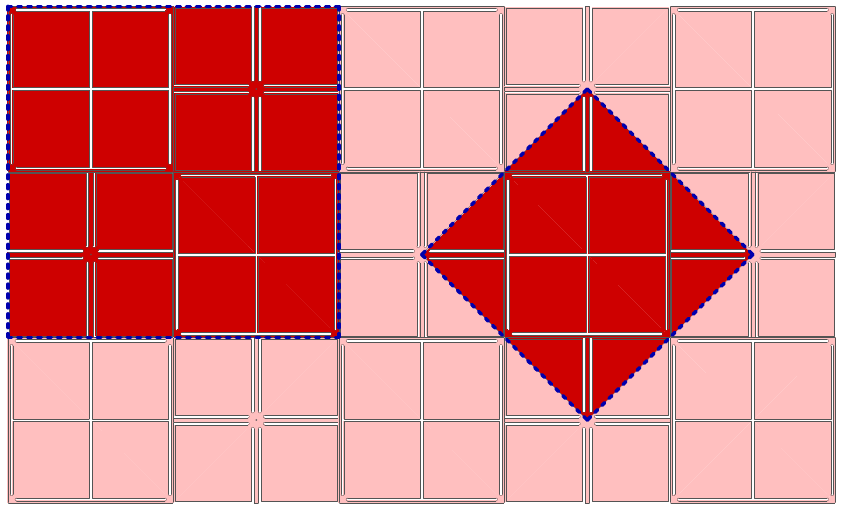}
	\includegraphics[width=0.4825\textwidth]{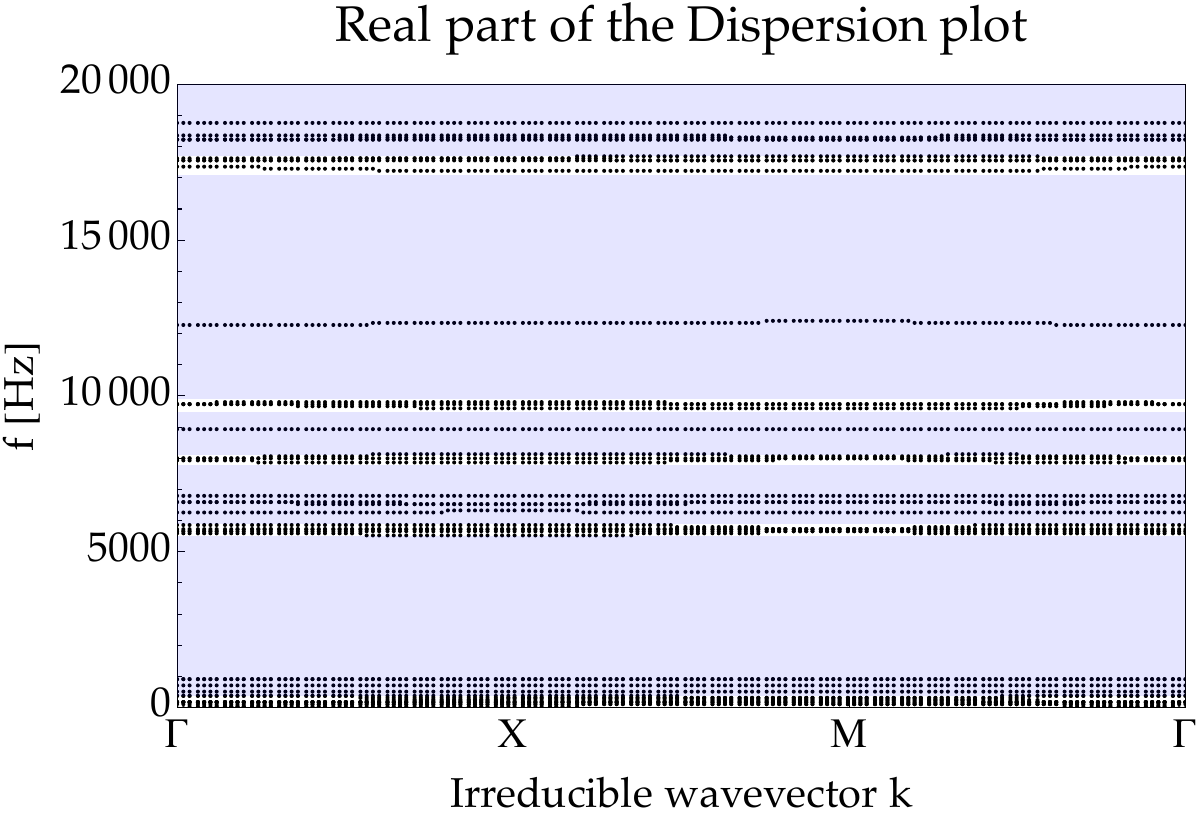}
	\includegraphics[width=0.48\textwidth]{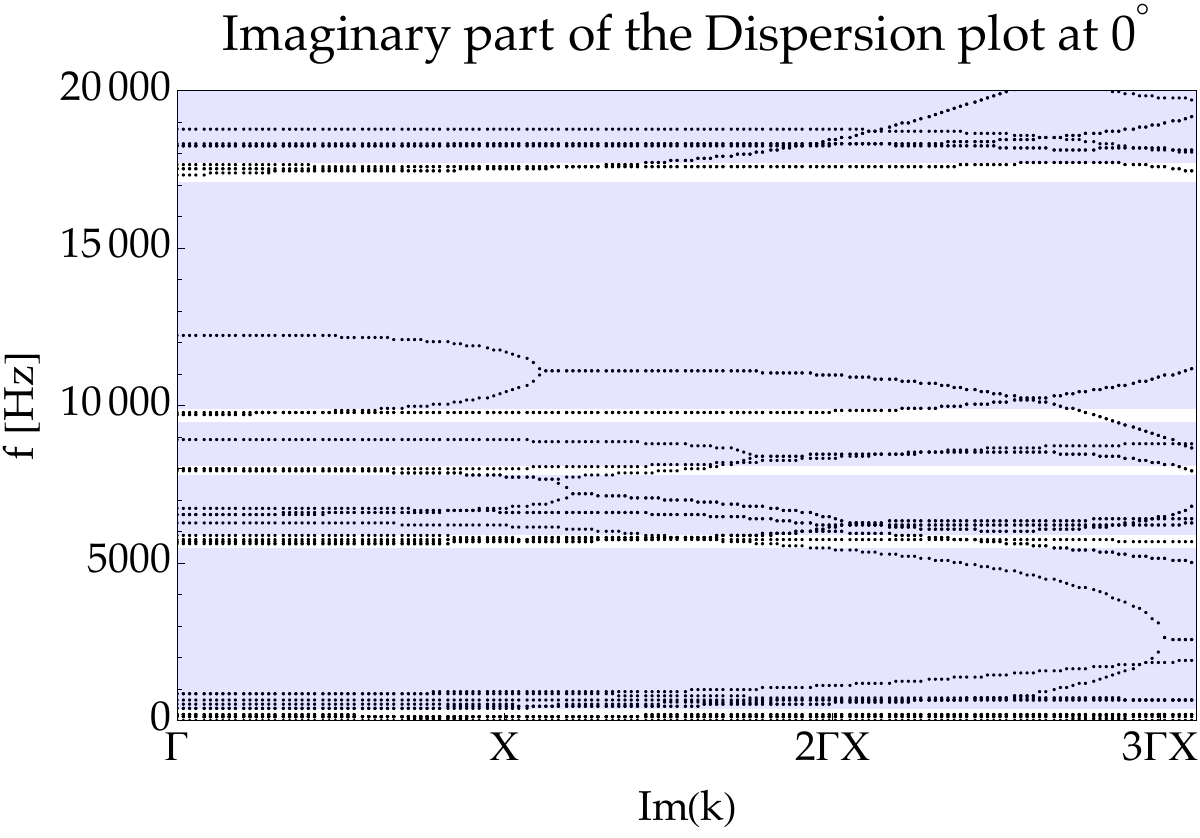}
	\caption{(Top) the FR $p4mm$ design that results by applying the dual cell method on unit cells $A$ and $B$  of Fig.~\ref{fig:FR_unit_cell} using a chessboard configuration. Two unit cells are highlighted with dark red. The first (left) is a conventional unit cell that shows the $AB$ coupling in a clear way while the second (right) is a primitive unit cell that also carries the full symmetry of the $p4mm$ wallpaper group. (Bottom) corresponding band strcuture with (left) real part and (right) imaginary part for wave propagation along $0^{\circ} $. In the imaginary part of the plot the wavevector used reaches up to three times the irreducible wavevector's maximum size. Omnidirectional band-gaps are highlighted in blue.}
	\label{fig:FR_chess_UC_DCs}
\end{figure}

\begin{figure}[h!]
	\centering
	\includegraphics[width=0.4825\textwidth]{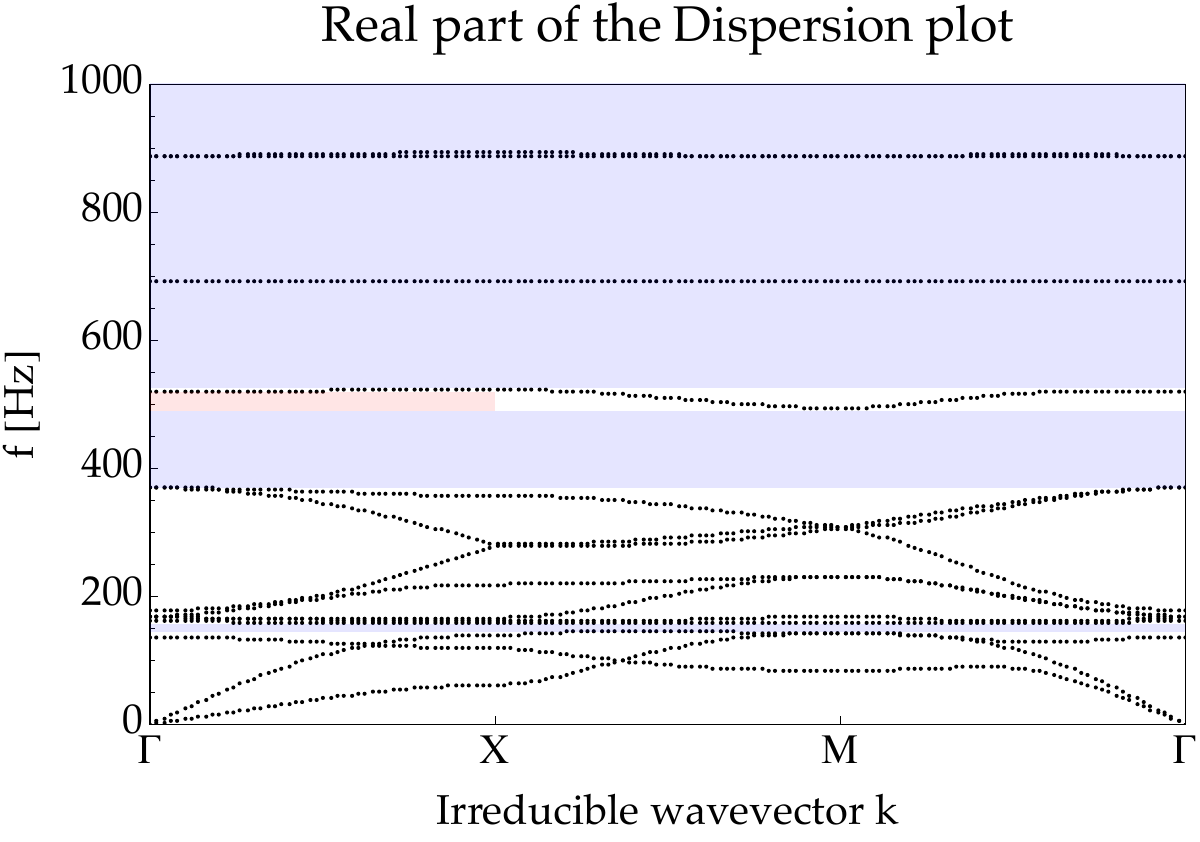}
	\includegraphics[width=0.48\textwidth]{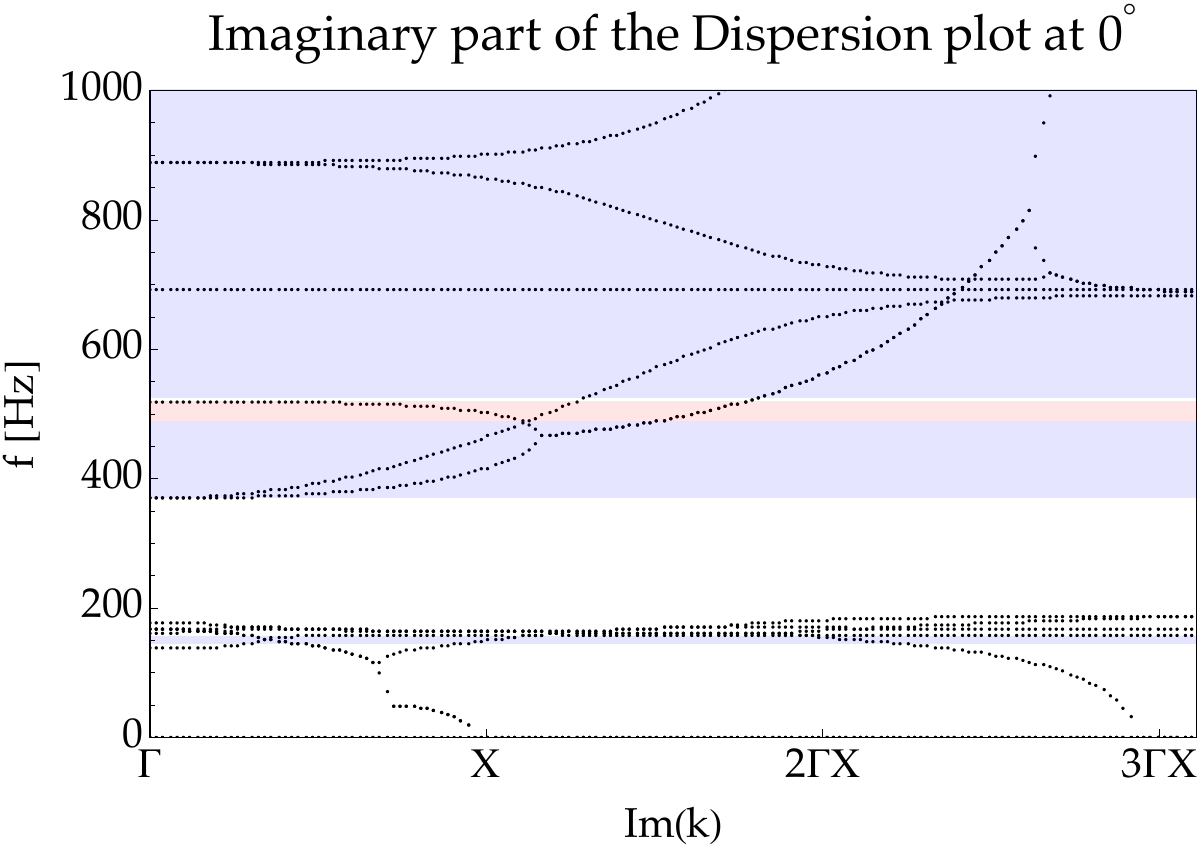}
	\caption{Band strcuture at low frequencies for the FR chessboard p4mm design of Fig.~\ref{fig:FR_chess_UC_DCs} with (left) real part and (right) imaginary part for wave propagation along $0^{\circ} $. In the imaginary part of the plot the wavevector used reaches up to three times the irreducible wavevector's maximum size. Omnidirectional band-gaps are highlighted in blue and directional band-gaps in red.}
	\label{fig:FR_chess_DCs_low}
\end{figure}

%
The chessboard FR $p4mm$ metamaterial that results from applying the dual cell method and a chessboard configuration can be seen in Fig.~\ref{fig:FR_chess_UC_DCs} (top) with two choices of unit cell highlighted with dark red inside blue dashed-line squares.
On the left, a conventional unit cell that shows the coupling in a clear way can be seen and on the right a primitive unit cell with the full symmetry of the $p4mm$ wallpaper group. 
The corresponding band structure can be seen at the bottom. 
Additionally, a detail of the band structure at low frequencies can be seen in Fig.~\ref{fig:FR_chess_DCs_low}.
Bloch-Floquet analysis that produces this band structure is done on the primitve unit cell shown, but rotated by $45^{\circ} $. 
\subsubsection{Discussion}\label{sec:chess_discuss}
From Figs.~\ref{fig:SCH_chess_UC_DCs} and~\ref{fig:FR_chess_UC_DCs} it is evident that the chessboard configuration is superior to the side-by-side in terms of omnidirectionality of band-gaps.
The FR chessboard $p4mm$ design is superior to the corresponding SCH chessboard  $p4mm$ design, specifically in the low frequency range, which is something anticipated given that the FR design takes advantage of the Local Resonance mechanism in addition to Bragg Scattering.
Specifically, the FR chessboard $p4mm$ design exhibits a very small omnidirectional band-gap at low frequencies (145-157 Hz) and then multiple omnidirectional band-gaps appear inbetween flat bands, spanning the frequency range from 370 Hz up to the end of the acoustic range with the exception of a small range from 490-525 Hz.
Given this results, the dual cell method  applied in 2D using a chessboard configuration seems to be a very promising strategy for broadband vibration mitigation, specifically for parent metamaterials of the Local Resonance type and calls for an extension of the method to 3D for more realistic applications.

\subsection{A finite-sized test for the FR chessboard $p4mm$ metamaterial}
In this section a finite-sized numerical test is presented for the best performing metamaterial of this study, the FR chessboard $p4mm$ design of Fig.~\ref{fig:FR_chess_UC_DCs}.
This test serves as an extra numerical verification of a correctly implemented Bloch-Floquet analysis.
For this test, a frequency domain study is performed using  \comsol, in order to obtain the Transmissibility plot for an $8 \times 8$ metamaterial, assessing its wave attenuation capabilities in the band-gap. 

\begin{figure}[!htbp]
	\centering
	\includegraphics[width=0.48\textwidth]{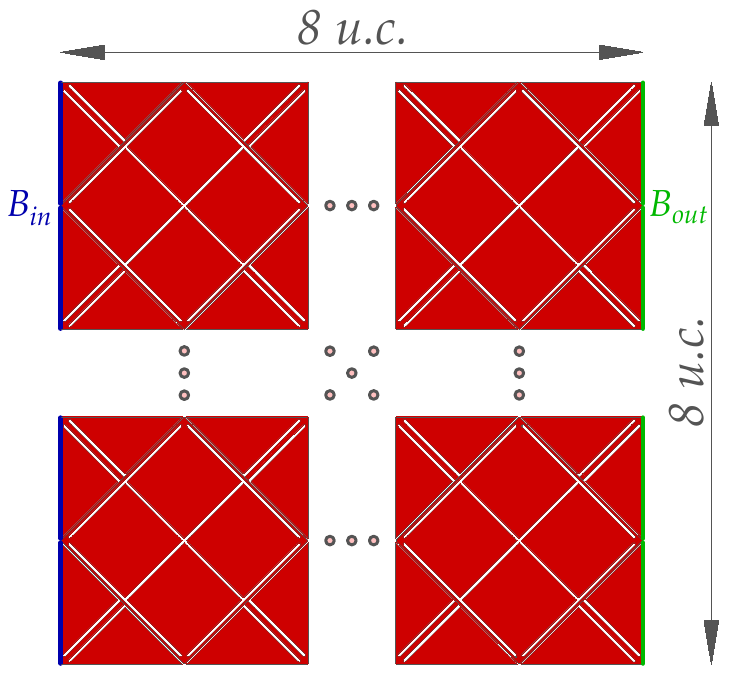}
	\caption{Shematic view of the test on the $8 \times 8$ FR chessboard $p4mm$ metamaterial. The input boundary $B_{in}$ is colored in blue while the output boundary $B_{out}$ is colored in green. Traction-free boundaries are colored in black.}
	\label{fig:chess_FR_transm_sch_view}
\end{figure}

A schematic view of the test can be seen in Fig.~\ref{fig:chess_FR_transm_sch_view}.
The excitation is a displacement applied on the input boundary $B_{in}$, which is colored with blue, and it is of the form
\begin{equation}
	\overline{u}_{\rm in}=u_{0}\cdot \hat{e}_{1} + 0\cdot \hat{e}_{2},
	\label{eq:inp_disp}
\end{equation}
where $\hat{e}_{1}$ and $\hat{e}_{2}$ are the unit vectors in the x- and y-direction respectively and $u_{0}=\SI{1}{\milli\meter}$.
The output boundary $B_{out}$ is colored with green and all the rest boundaries are traction-free and colored with black. 

We define Transmissibilty (\textit{T}) as the ratio of input dislplacement to output displacement. 
More specifically
\begin{equation}
	T=\dfrac{
		\dfrac{1}{\left| B_{\rm out} \right|}\int\limits_{B_{\rm out}} \sqrt{\left| u_1 \right|^2 }
	}{
		\dfrac{1}{\left| B_{\rm in} \right|}\int\limits_{B_{\rm in}} \sqrt{\left| \overline{u}_1 \right|^2}
	} \, ,
	\label{eq:transmission}
\end{equation}
where $u_1$ is the x-component of the displacement field ($u$), $\overline{u}_1$ is the x-component of the prescribed input displacement which is applied on the input boundary ($\overline{u}_{\rm in}$), and $B_{\rm in}$ and $B_{\rm out}$ denote the input and output boundaries, respectively.
Since the analysis is performed in the frequency domain, the displacement components are complex, which means that taking the absolute value  of a component gives its magnitude, corresponding to the maximum amplitude at a given frequency. 

\begin{figure}[!htbp]
	\centering
	\includegraphics[width=0.48\textwidth]{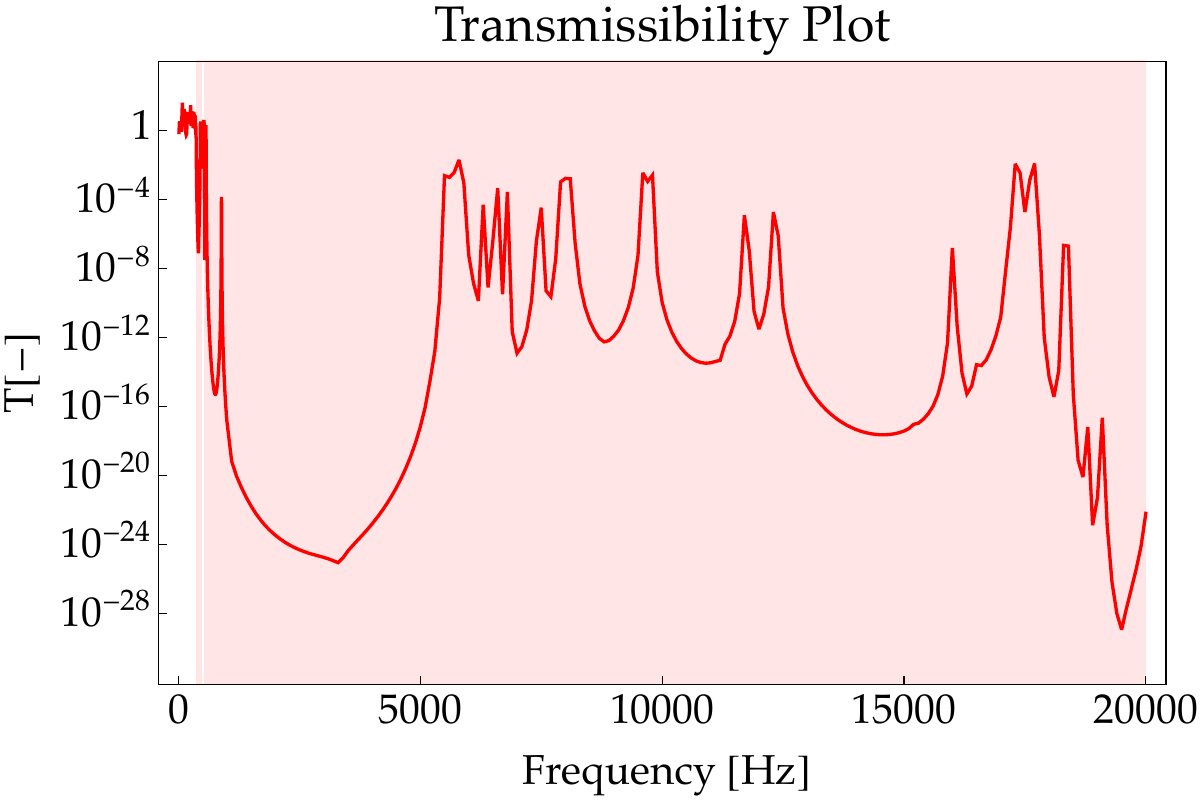}
	\includegraphics[width=0.48\textwidth]{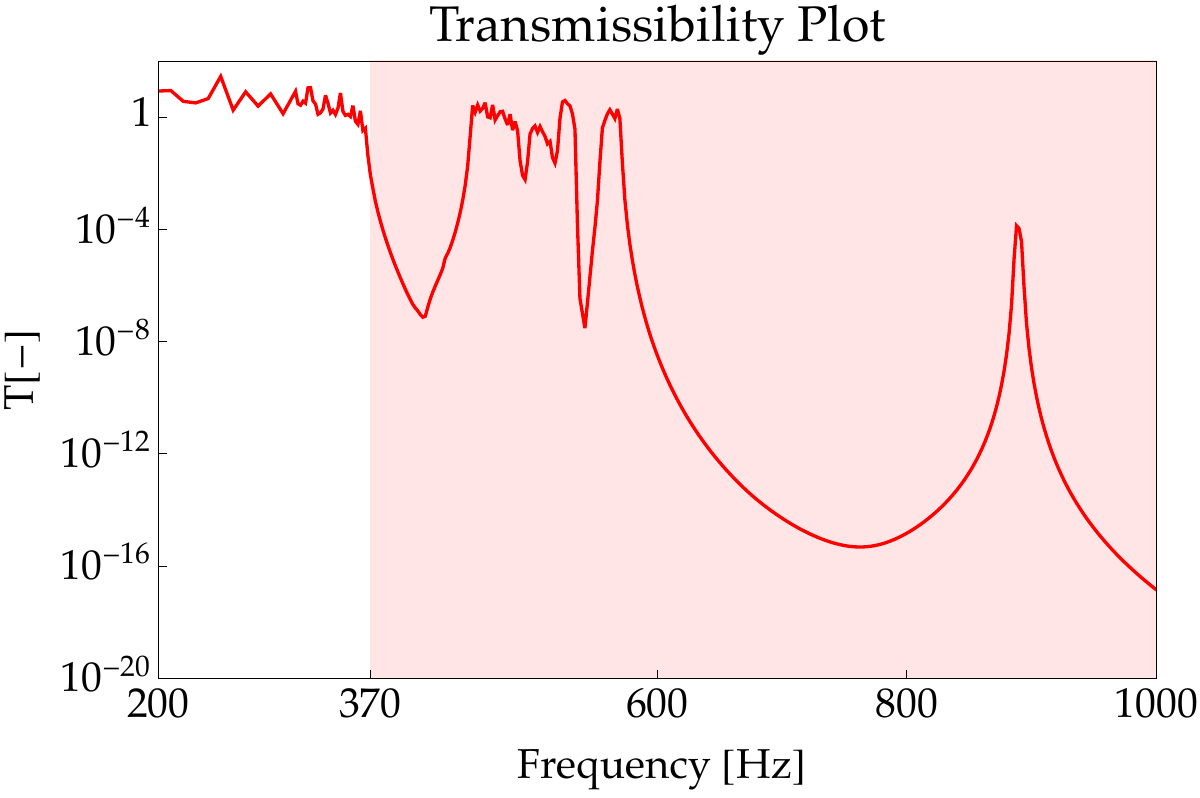}
	\caption{Transmissibility plot for an $8\times 8$ FR chessboard $p4mm$ metamaterial: (left) the full plot (right) the low frequency range. The band-gap range is highlighted in red and the y-axis uses a logarithmic scale.}
	\label{fig:chess_FR_transm}
\end{figure}

The input displacement's form in Eq.~(\ref{eq:inp_disp}) shows that the test simulates a pressure wave travelling in the x-direction. 
Simulations are carried out with the ouput mesh of a mesh convergence study and the Transmissibility plot is shown in Fig.~\ref{fig:chess_FR_transm} with (left) the full plot up to the end of the acoustic range and (right) a part of the plot emphasizing the low frequency range.
The Transmissibility plot confirms the band-structure calculations and the abilty of the dual cell method to produce multiple effective band-gaps, as many consecutive band-gaps are evident,  some of which also exhibit strong attenuation performance.
%
\section{Conclusions and future perspective}\label{sec:concl}
%
The results of this work and in particular those of section~\ref{sec:chess} have some particularly interesting implications for the design of metamaterials.
Firstly, results support the idea that any band-gap metamaterial could have a much more efficient version which can be constructed purely from its own unit cells, at the cost of enlarging the lattice constant by a certain factor.
In 2D, this factor has a value of $\sqrt{2}$ for square unit cells given that the chessboard pattern of two different square unit cells has a primitive cell which is again a square, but with a side length bigger by a $\sqrt{2}$ factor and rotated by $45 ^{\circ}$, as can be seen in Figs.~\ref{fig:SCH_chess_UC_DCs} and~\ref{fig:FR_chess_UC_DCs}.

\begin{figure}[h!]
	\centering
	\includegraphics[width=0.495\textwidth]{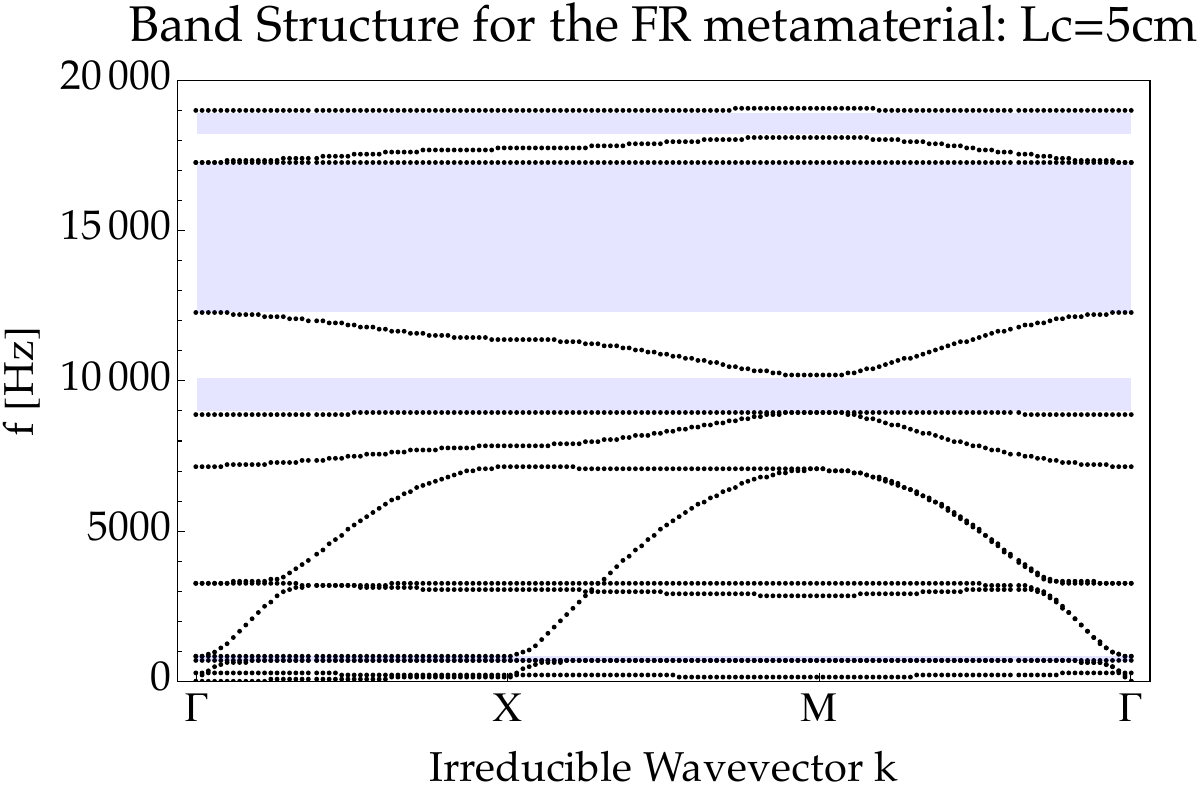}
	\includegraphics[width=0.48\textwidth]{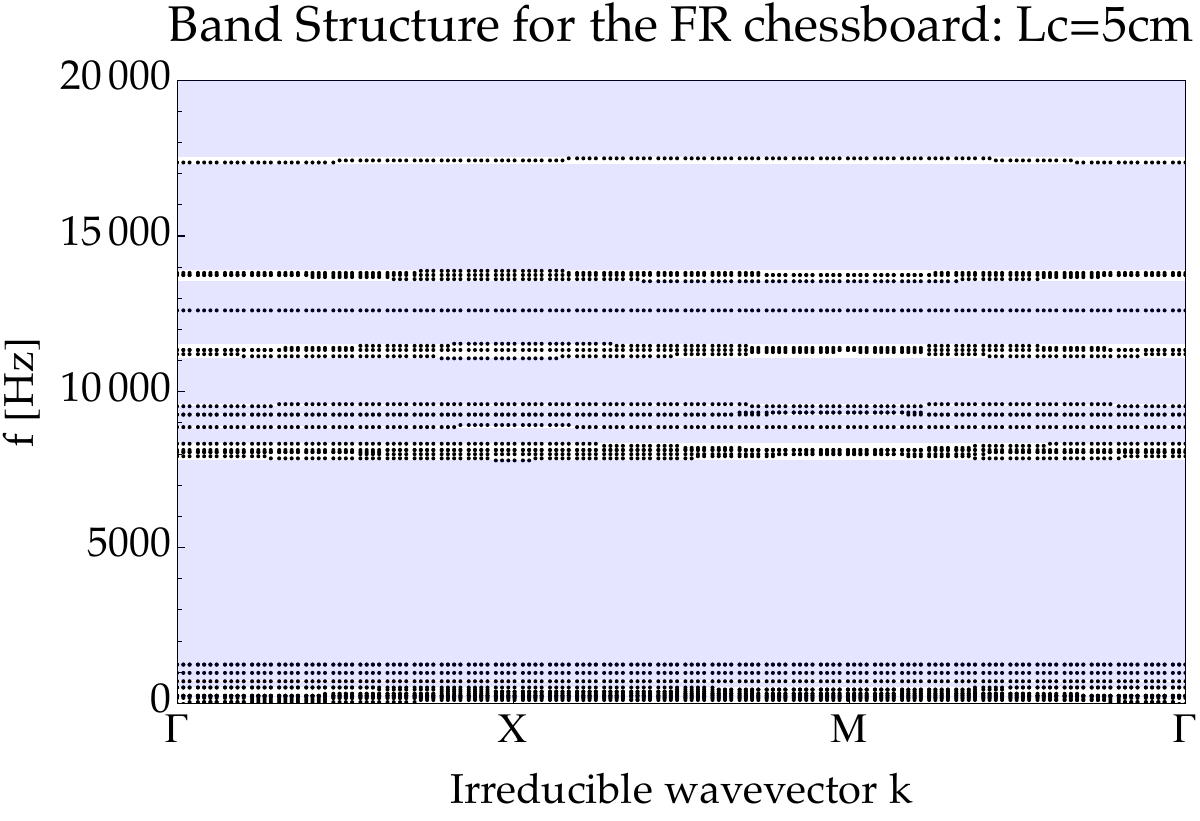}
	\includegraphics[width=0.47\textwidth]{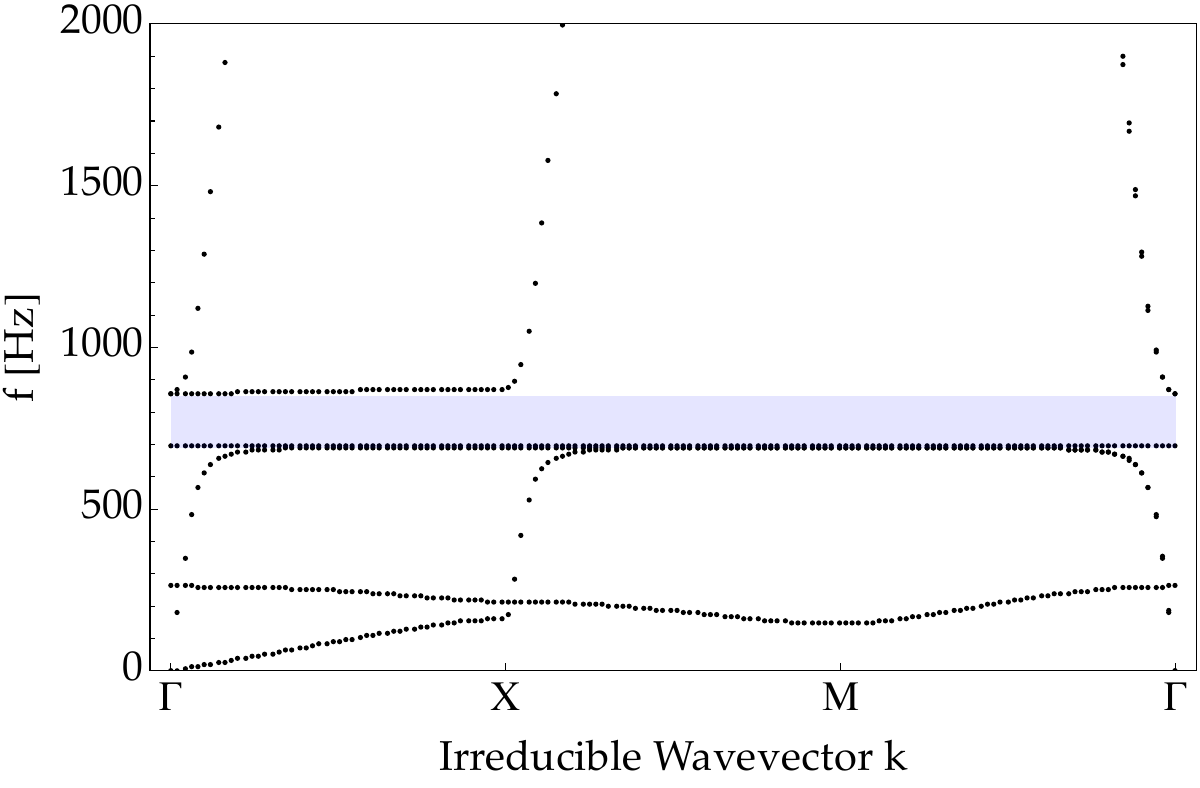}
	\includegraphics[width=0.4735\textwidth]{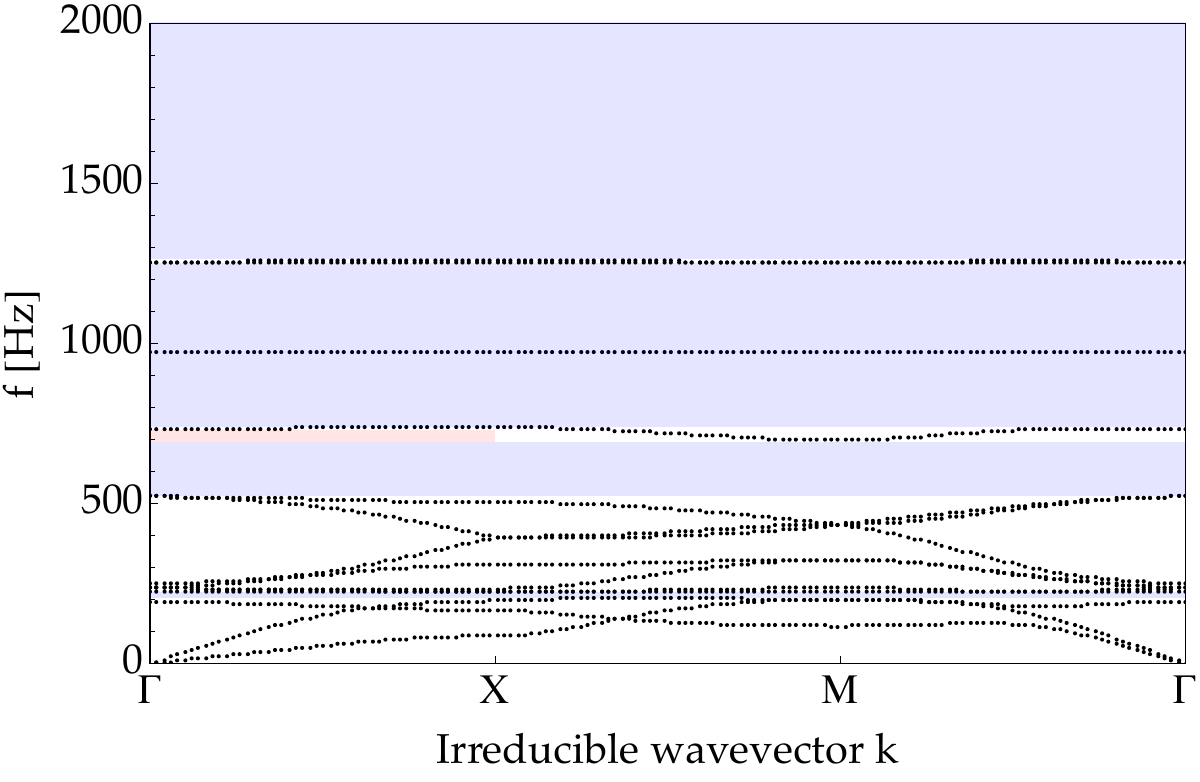}
	\caption{Copmarison of the band structure for the FR metamaterial with side length $L_{c}=\SI{5}{\centi\meter}$ and the FR chessboard metamaterial with equal side length. Band structure is presented for (left) the FR metamaterial and (right) the FR chessboard metamaterial, with band structures up to (top) the end of the acoustic range and (bottom) $\SI{2000}{\hertz}$. }
	\label{fig:FR_parent_chess_comp}
\end{figure}

Secondly, if for a certain application a square unit cell has already been designed, optimized and its size must remain constant (enlarging the lattice constant cannot be tolerated), then the unit cell can be scaled down by a factor of $\frac{1}{\sqrt{2}}$, and the dual cell method using the chessboard configuration can be applied, producing a primitive cell with the same size as the original cell, exhibiting additional more efficient band-gaps. 
One should note that in this way the lowest band-gap inherited by the parent metamaterial is shifted higher, but it is still possible that it is replaced by another gap and in any case the higher frequency range will be optimised. 
This is very close to what happens when the FR parent metamaterial is compared to the FR chessboard $p4mm$ that results from this procedure and therefore has the same side length $L_{c}$, as can be seen in Fig.~\ref{fig:FR_parent_chess_comp}. 
The Local Resonance gap that had a starting frequency around $\SI{700}{\hertz}$  for the FR parent metamaterial has moved to a higher frequency range in the FR chessboard design due to the shrinking of the resonators by the factor $\frac{1}{\sqrt{2}}$.
However, we witness that new Local Resonance gaps appear at the frequency range where we had the Local Resonance gap in the parent metamaterial (except from the range $700-740~ \text{Hz}$ where only a directional band-gap exists), and additional Bragg-gaps appear even lower, while the higher frequency range has been enhanced with multiple additional Bragg gaps\footnote{whether the band-gaps for the chessboard metamaterial are Bragg or Local Resonance type can be understood by comparison of its band structure in Fig.~\ref{fig:FR_parent_chess_comp} with the imaginary part of the band structure on Figs.~\ref{fig:FR_chess_UC_DCs} and~\ref{fig:FR_chess_DCs_low}, as a linear scaling of the diseprsion plot happens when only the size of the unit cell is changed.}, which shows that the dual cell method has produced a much more efficient metamaterial without the cost of enlarging the lattice constant in this case.
It must be noted that the ``replacement" of the Local Resonance gap with other gaps indeed happened in this one particular case but only partly due to the $\SI{40}{\hertz}$ range where only the directional band-gap exists. 
Thus a generalisation of this phenomenon is not suggested yet for every metamaterial design, as it is highly likely dependent on the coupling configuration of the two unit cells of the parent metamaterial chosen and the unit cell choices themselves. 

Furthermore, the multiple consecutive band-gaps that appeared in-between flat-bands throughout this work as a result of the dual cell method, some of which also provide enhanced attenuation,  show that the dual cell method looks promising in providing broadband vibration absorption to already optimised metamaterial designs, acting as an ``add-on" to structural optimisation.

For the future, the dual cell method could be applied to metasurfaces and can be extended to 3D with many potential applications in bulk mechanical and acoustic metamaterials. 
The method can also be extended to mechanical metamaterials that exhibit band-gaps due to Inertial Amplification, but due to the geometries of proposed metamaterial inerters, it is possible that the coupling of two unit cells becomes challenging and may require coupling elements.

Lastly, given that the two band-gap mechanisms exploited in this work have an electromagnetic analogue, the dual cell method can potentially also be applied to photonic crystals and electromagnetic metasurfaces with many potential applications.

\vspace{10pt}
\begingroup
\scriptsize
\noindent
\textbf{Acknowledgements.}
Plastiras Demetriou acknowledges support from Gianluca Rizzi for his valuable feedback and insightful comments on parts of this work.
\\
\noindent
\textbf{Conflict of interest.} The author declares that they have no conflicts of interest in this work.
\endgroup
\printbibliography
\appendix
\section{1-D Spring-mass chains: Equilibrium Equations in Matrix Form}\label{sec:eq_eq}
In this section, the Equilibrium Equations for each chain are presented in the form of a linear homogeneous system. 
For any spring-mass chain, one can arrive at the Equilibrium Equations by using Newton's law or Lagrangian Mechanics \cite{brillouin1953wave}, \cite{xu2021hybrid}, \cite{dove2011introduction}.
Starting from these equations, the Bloch-Floquet theorem is used and a plane wave ansatz is assumed for each displacement, with the corresponding wave amplitude named with the corresponding displacement's capital lettter.
For example, for the displacement $u_{n}$ of a mass $m$ at the $n_{th}$ unit cell we have:
\begin{equation}
	u_{n}=Ue^{i(kna-\omega t)},
	\label{plane_wave}
\end{equation}
where $U$ is the corresponding wave amplitude, $k$ is the wavenumber, $\omega$ is the angular frequency, $t$ is time and $a$ the lattice constant.\\
Then, the system can be rearranged in a matrix form which is presented in the following.
In this way, the so-called Acoustic Tensor appears clearly and one can calculate the dispersion polynomial by simply setting its determinant to zero.  For the dispersion relations one can either solve for the angular frequency $\omega$ or wavenumber $k$ which are the two unknowns.

\subsection{The diatomic chain of Fig.~\ref{fig:diatomic}}
For the diatomic chain we have two degrees of freedom, namely: $u$ and $v$, for the displacement of mass 1($m_{1}$) and mass 2 ($m_2$) respectively.
The linear homogenous system reads:
\begin{equation}
	\begin{bmatrix}
		2k_{1} - m_{1}\omega^{2}  & -k_{1}(1 + e^{-ika})\\
		-k_{1}(1 + e^{ika}) & 2k_{1} - m_{2}\omega^{2} 
	\end{bmatrix}
	\begin{bmatrix}
		U\\
		V  
	\end{bmatrix}
	=
	\begin{bmatrix}
		0 \\
		0 
	\end{bmatrix},
\end{equation}
where the leftmost matrix is the Acoustic Tensor, and the values of the parameters are given in Table~\ref{tab:diatomic_properties}.
\begin{table}[h]
	\centering
	\begin{tabular}{|l|c|c|}
		\hline
		\textbf{Parameter} & \textbf{Symbol} & \textbf{Value} \\
		\hline
		mass 1 & $m_{1}$ & \SI{1}{\kilogram} \\
		\hline
		mass 2       & $m_{2}$ & \SI{0.1}{\kilogram}\\
		\hline
		spring stiffness              & $k_{1}$ & \SI{1}{\newton\per\meter} \\
		\hline
		lattice constant     & a & \SI{2}{\meter}\\
		\hline
	\end{tabular}
	\caption{Parameter values for the diatomic chain}
	\label{tab:diatomic_properties}
\end{table}
\subsection{The tetra-atomic chain of Fig.~\ref{fig:tetraatomic} resulting from the dual cell method}
For the tetra-atomic chain  we have four degrees of freedom, namely: $u, v ,w$ and $x$, for the displacement of each mass of the dual cell in Fig.~\ref{fig:tetraatomic} respectively, starting from the leftmost mass and moving towards the right. 
The linear homogeneous system reads:
\begin{equation}
	\begin{bmatrix}
		2k_{1} - m_{1}\omega^{2}  & -k_{1} & 0 & -k_{1}e^{-ika}\\
		-k_{1}& 2k_{1} - m_{2}\omega^{2}  & -k_{1} & 0 \\
		0 & -k_{1} & 2k_{1} -m_{2}  \omega^{2} & -k_{1}\\
		-k_{1}e^{ika} & 0 & -k_{1} & 2k_{1} - m_{1}\omega^{2}  \\
	\end{bmatrix}
	\begin{bmatrix}
		U\\
		V \\
		W\\
		X
	\end{bmatrix}
	=
	\begin{bmatrix}
		0 \\
		0 \\
		0\\
		0
	\end{bmatrix},
\end{equation}
where the leftmost matrix is the Acoustic Tensor, the values of the masses and the spring constant are the same as in Table~\ref{tab:diatomic_properties} and the lattice constant is effectively doubled due to the dual cell method.
\subsection{The mass-in-mass (Locally resonant chain) of Fig.~\ref{fig:mass_in_mass}}
For the mass-in-mass chain we have two degrees of freedom, namely $u$ and $v$ for the displacement of the primary mass ($M$) and the displacement of the resonator mass ($m$) respectively. 
The linear homogeneous system reads: 
\begin{equation}
	\begin{bmatrix}
		2K_{1} (1-cos(ka)) + k_{1} - M\omega^{2}& -k_{1}\\
		-k_{1} & k_{1}-m\omega^{2}
	\end{bmatrix}
	\begin{bmatrix}
		U\\
		V  
	\end{bmatrix}
	=
	\begin{bmatrix}
		0 \\
		0 
	\end{bmatrix},
\end{equation}
where the leftmost matrix is the Acoustic Tensor and the parameter values are given in Table~\ref{tab:mass_in_mass_properties}.
\begin{table}[!htbp]
	\centering
	\begin{tabular}{|l|c|c|}
		\hline
		\textbf{Parameter} & \textbf{Symbol} & \textbf{Value} \\
		\hline
		primary mass& $M$ & \SI{2}{\kilogram} \\
		\hline
		resonator mass       & $m$ & \SI{1.2}{\kilogram}\\
		\hline
		primary spring stiffness              & $K_{1}$ & \SI{2}{\newton\per\meter} \\
		\hline
		resonator spring stiffness              & $k_{1}$ & \SI{0.3}{\newton\per\meter} \\
		\hline
		lattice constant     & a & \SI{2}{\meter}\\
		\hline
	\end{tabular}
	\caption{Parameter values for the mass-in-mass chain}
	\label{tab:mass_in_mass_properties}
\end{table}
\subsection{The hybrid triatomic mass-in-mass chain of Fig.~\ref{fig:dual_from_mass_in_mass} resulting from the dual cell method}
For the hybrid triatomic mass-in-mass chain we have six degrees of freedom, namely: $u, v, w, x, y$ and $z$, where $u, v$ and $w$ are the displacements of primary masses $M_{1}, M_{2}$ and $M_{3}$ respectively (going from left to right in the unit cell shown in Fig.~\ref{fig:dual_from_mass_in_mass}) and $x, y$ and $z$ are the displacements of their resonator masses ($m_i$) respectively. 
The values of all the parameters are reported in Table~\ref{tab:tri_mass_in_mass_properties} for clarity.
The indexing of the resonator masses ($m_i$) and resonator springs  ($k_i$)  follows the same convention with the primary masses  ($M_i$) , while the indexing of the primary springs ($K_i$) is the only one differing: primary spring 1 ($K_1$) is the one after primary mass 1 ($M_1$) , primary spring 2 ($K_2$) is the one after primary mass 2 ($M_2$) and primary spring 3 ($K_3$)  is the one after primary mass 3 ($M_3$) , which implies that the first primary spring inside the unit cell in Fig.~\ref{fig:dual_from_mass_in_mass} is primary spring 3 ($K_3$).
The linear homogeneous system reads:
\begin{equation}
	\resizebox{\textwidth}{!}{%
		$\begin{bmatrix}
			2K_{1}+k_{1} - M_{1}\omega^{2}  & -K_{1}  & -K_{1}e^{-ika}& -k_{1} & 0&0\\
			-K_{1}& K_{1}+K_{2}+k_{2} - M_{2}\omega^{2}  & -K_{2} & 0& -k_{2} & 0 \\
			-K_{1}e^{ika} & -K_{2} & K_{2}+K_{1}+k_{2} - M_{2}\omega^{2} & 0 & 0 &-k_{2}\\
			-k_{1} & 0 & 0& k_{1}-m_{1}\omega^{2} & 0 & 0\\
			0 & -k_{2} & 0& 0& k_{2}-m_{2}\omega^{2} & 0\\
			0 & 0& -k_{2}& 0& 0 &k _{2}- m_{2}\omega^{2}\\
		\end{bmatrix}
		\begin{bmatrix}
			U\\
			V \\
			W\\
			X\\
			Y\\
			Z
		\end{bmatrix}
		=
		\begin{bmatrix}
			0 \\
			0 \\
			0\\
			0\\
			0\\
			0\\
		\end{bmatrix},$
	}
\end{equation}
where the leftmost matrix is the Acoustic Tensor.
\begin{table}[!htbp]
	\centering
	\begin{tabular}{|l|c|c|}
		\hline
		\textbf{Parameter} & \textbf{Symbol} & \textbf{Value} \\
		\hline
		primary mass 1       & $M_{1}$ & \SI{2}{\kilogram} \\
		\hline
		primary mass 2       & $M_{2}$ & \SI{1}{\kilogram}\\
		\hline
		primary mass 3      & $M_{1}$ & \SI{1}{\kilogram} \\
		\hline
		resonator mass 1       & $m_{1}$ & \SI{1.2}{\kilogram} \\
		\hline
		resonator mass 2       & $m_{2}$ & \SI{0.6}{\kilogram}\\
		\hline
		resonator mass 3      & $m_{2}$ & \SI{0.6}{\kilogram} \\
		\hline
		primary spring 1 stiffness              & $K_{1}$ & \SI{4}{\newton\per\meter} \\
		\hline
		primary spring 2 stiffness              & $K_{2}$ & \SI{2}{\newton\per\meter} \\
		\hline
		primary spring 3 stiffness              & $K_{1}$ & \SI{4}{\newton\per\meter} \\
		\hline
		resonator spring 1 stiffness               & $k_{1}$ & \SI{0.3}{\newton\per\meter} \\
		\hline
		resonator spring 2 stiffness              & $k_{2}$ & \SI{0.6}{\newton\per\meter} \\
		\hline
		resonator spring 3 stiffness            & $k_{2}$ & \SI{0.6}{\newton\per\meter} \\
		\hline
		lattice constant     & a & \SI{4}{\meter}\\
		\hline
	\end{tabular}
	\caption{Parameter values for the triatomic mass-in-mass chain}
	\label{tab:tri_mass_in_mass_properties}
\end{table}
\section{Band Structures for the rest of the p2mm configuration combinations}\label{sec:band_str_p2mm}

\subsection{SCH metamaterial p2mm designs}
\begin{figure}[!htbp]
	\centering
	\includegraphics[width=0.48\textwidth]{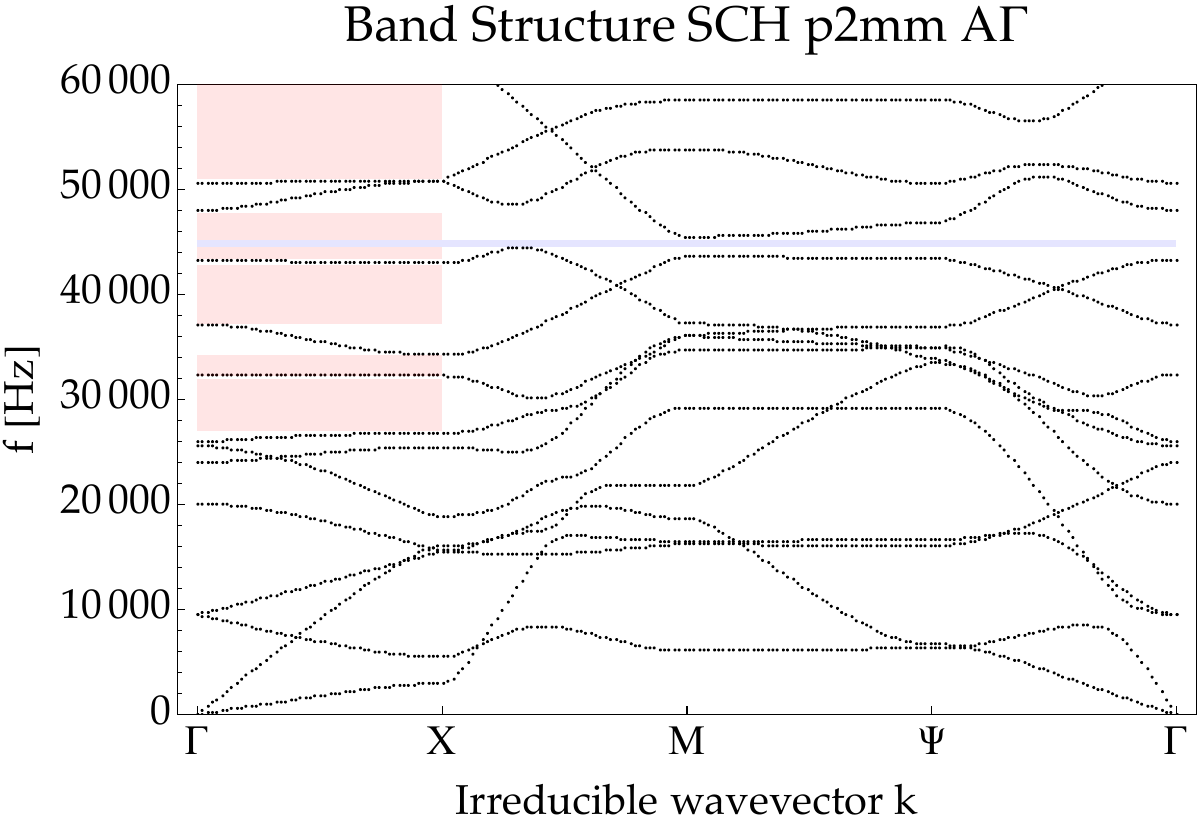}
	\includegraphics[width=0.48\textwidth]{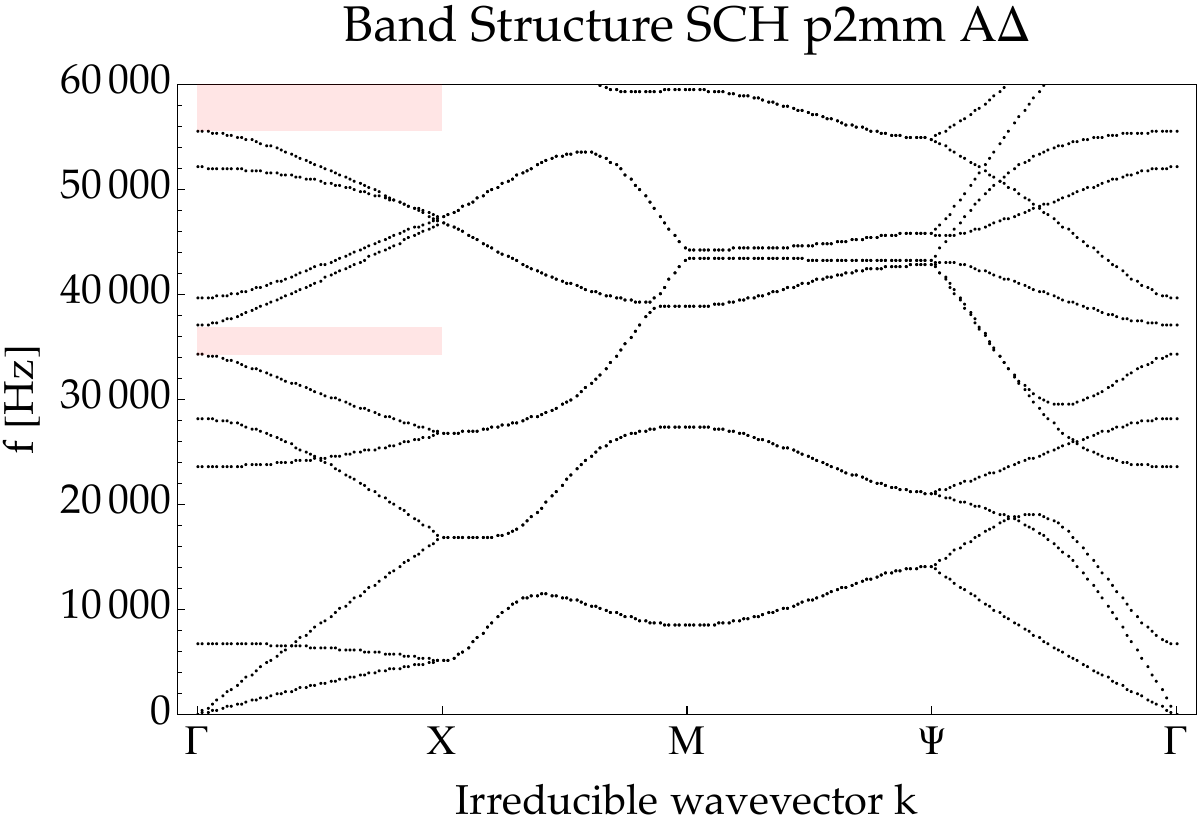}
	\caption{(Left) Band structure for the SCH p2mm design using unit cells $A$ and $\Gamma$ and (right) using unit cells $A$ and $\Delta$}
	\label{fig:dual_SCH_DC_p2mm_rest}
\end{figure}
\subsection{FR metamaterial p2mm designs}
\begin{figure}[!htbp]
	\centering
	\includegraphics[width=0.48\textwidth]{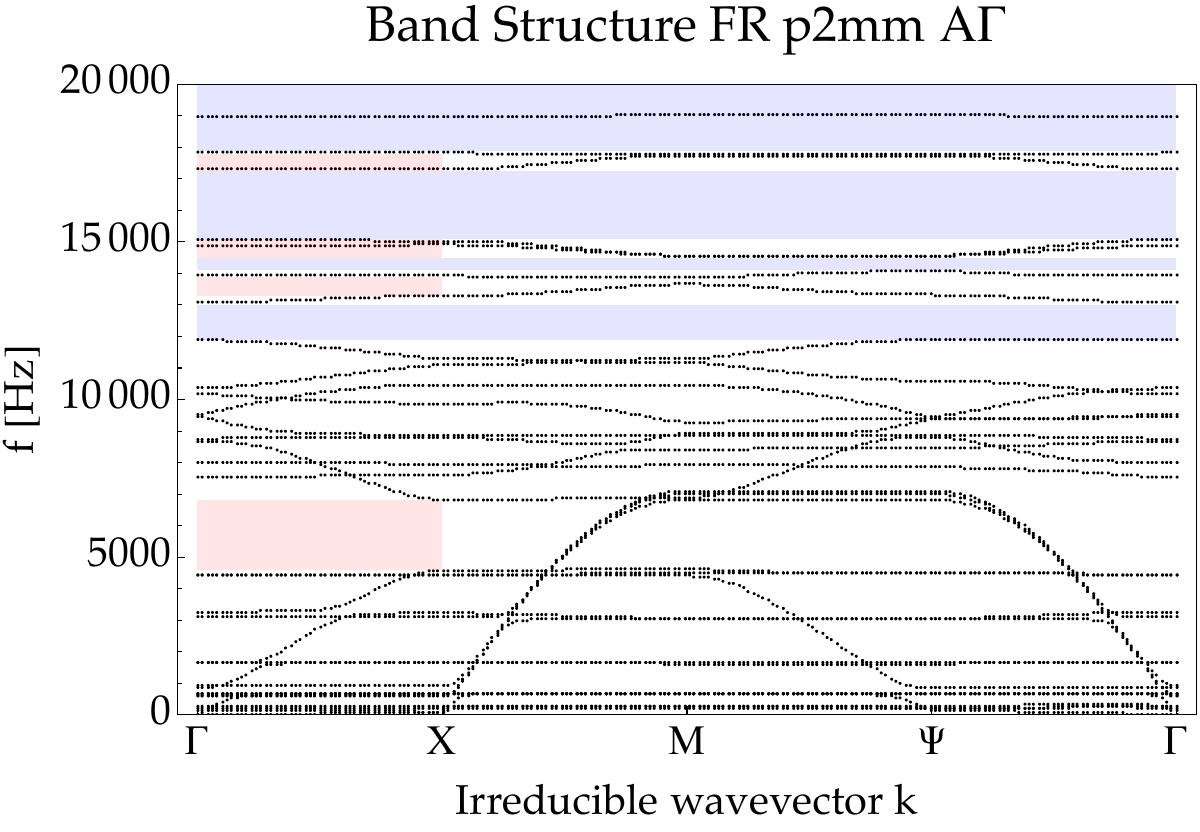}
	\includegraphics[width=0.48\textwidth]{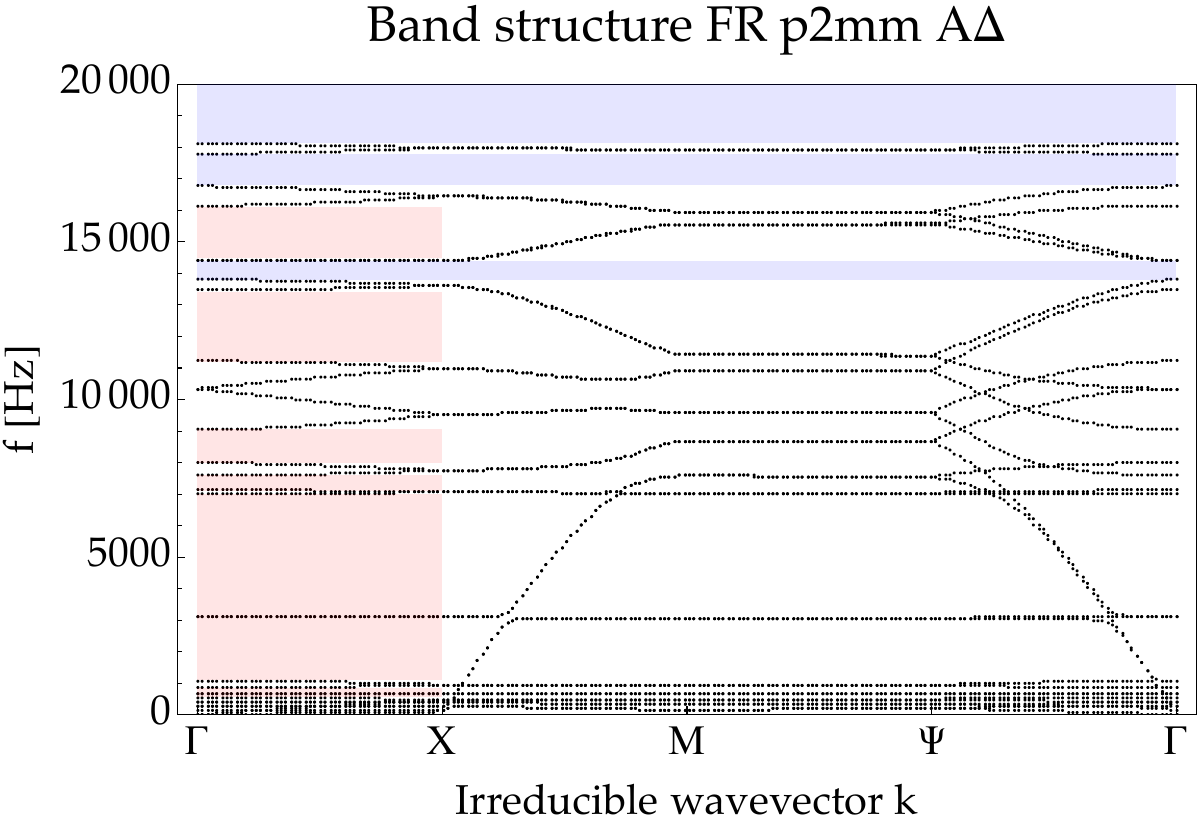}
	\caption{(Left) Band structure for the FR p2mm design using unit cells $A$ and $\Gamma$ and (right) using unit cells $A$ and $\Delta$}
	\label{fig:dual_FR_DC_p2mm_rest}
\end{figure}
\section{Band Structure for the FR metamaterial up to the end of the acoustic range}\label{sec:band_str_FR_app}
\begin{figure}[!htbp]
	\centering
	\includegraphics[width=0.48\textwidth]{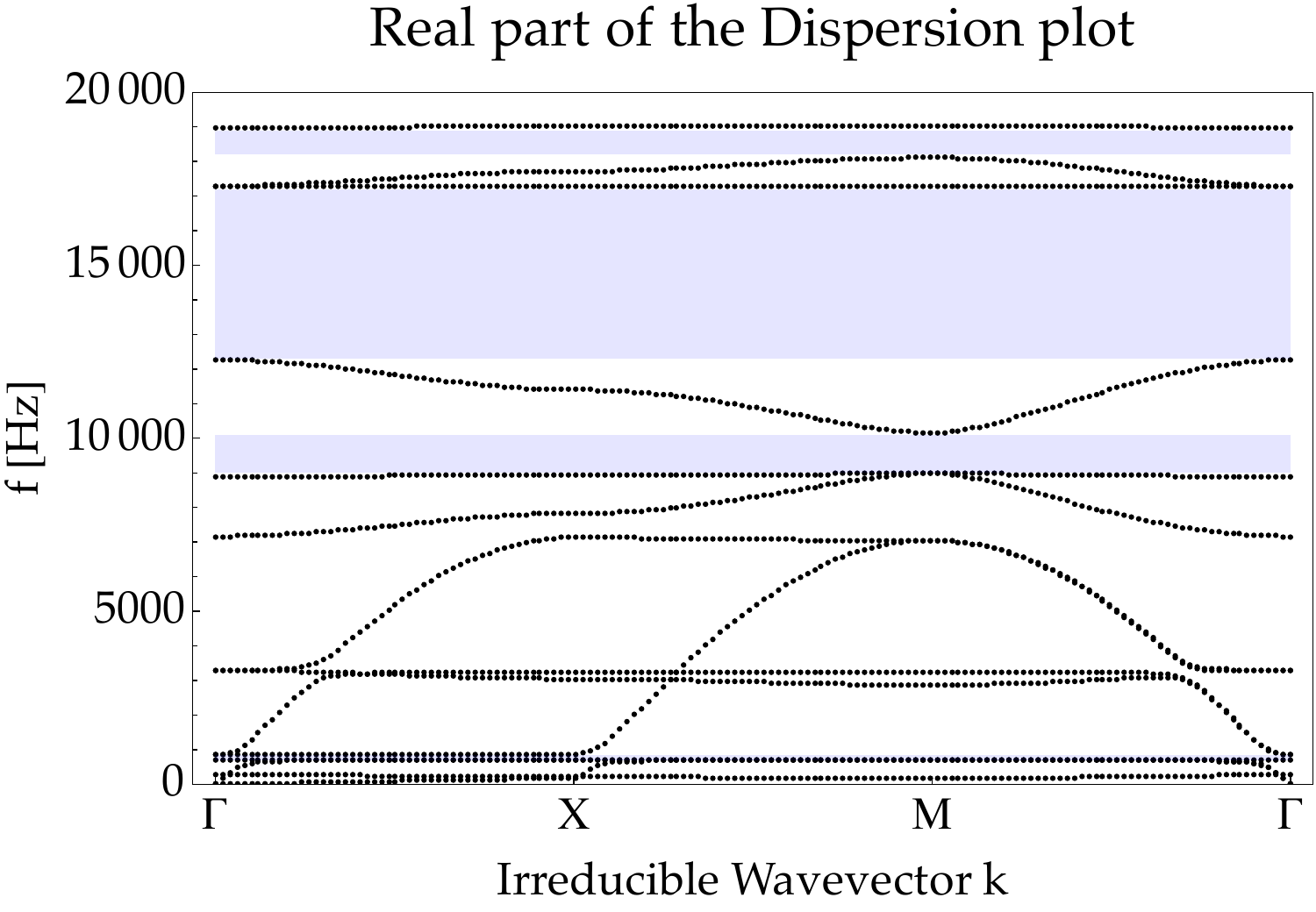}
	\includegraphics[width=0.48\textwidth]{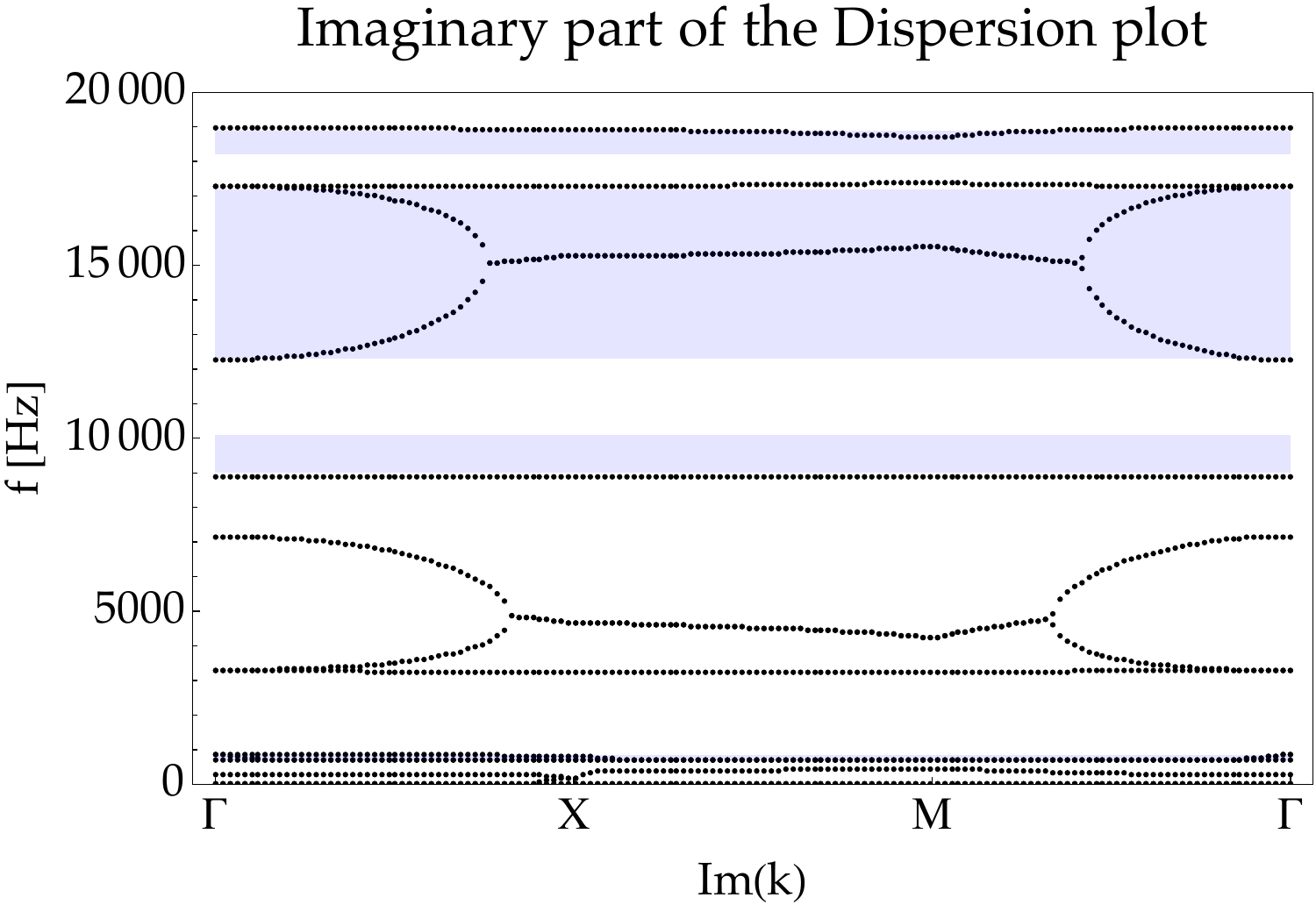}
	\caption{Band structure for the FR metamaterial up to the end of the acoustic range with (left) real part of the plot and (right) imaginary part of the plot. Omnidirectional band-gaps are highlighted in blue.}
	\label{fig:FR_unit_cell_DCs_full}
\end{figure}

\end{document}